\documentclass[fleqn,usenatbib]{mnras}

\usepackage{newtxtext,newtxmath}

\usepackage[T1]{fontenc}
\usepackage{ae,aecompl}

\usepackage{graphicx}	
\usepackage{amsmath}	
\usepackage{amssymb}	
\usepackage[utf8]{inputenc}
\usepackage{xcolor}

\newcommand{\kepler}{{\em Kepler}}

\newcommand{\cd}{\mbox{d$^{-1}$}}
\newcommand{\nurot}{\mbox{$\nu_{\rm rot}$}}

\newcommand{\muhz}{\mbox{$\muup$Hz}}

\title[\textit{Kepler} roAp stars]{Six new rapidly oscillating Ap stars in the \textit{Kepler} long-cadence data using super-Nyquist asteroseismology}

\author[D.~R.~Hey et al.]
{Daniel R. Hey,$^{1,2}$\thanks{E-mail: daniel.hey@sydney.edu.au}
Daniel L. Holdsworth,$^{3}$
Timothy R. Bedding,$^{1,2}$
\newauthor
Simon J. Murphy,$^{1,2}$
Margarida S. Cunha,$^{4}$
Donald W. Kurtz,$^{3}$
Daniel Huber,$^{5}$
\newauthor
Benjamin Fulton,$^{6}$ and
Andrew W.\ Howard$^{7}$
\\
$^{1}$School of Physics, Sydney Institute for Astronomy (SIfA), The University of Sydney, NSW 2006, Australia\\
$^{2}$Stellar Astrophysics Centre, Aarhus University, DK-8000 Aarhus C, Denmark \\
$^{3}$Jeremiah Horrocks Institute, University of Central Lancashire, Preston PR1 2HE, UK \\
$^{4}$Instituto de Astrof\'\i sica e Ci\^encias do Espa\c co, Universidade do Porto, CAUP, Rua das Estrelas, PT4150-762 Porto, Portugal\\
$^{5}$Institute for Astronomy, University of Hawai‘i, 2680 Woodlawn Drive, Honolulu, HI 96822, USA\\
$^{6}$NASA Exoplanet Science Institute / Caltech-IPAC, Pasadena, CA 91125, USA\\
$^{7}$California Institute of Technology, Pasadena, CA 91125, USA\
}

\date{Accepted XXX. Received YYY; in original form ZZZ}

\pubyear{2019}

\begin{document}
\label{firstpage}
\pagerange{\pageref{firstpage}--\pageref{lastpage}}
\maketitle

\begin{abstract}
We perform a search for rapidly oscillating Ap stars in the \textit{Kepler} long-cadence data, where true oscillations above the Nyquist limit of 283.21\,\muhz\ can be reliably distinguished from aliases as a consequence of the barycentric time corrections applied to the \textit{Kepler} data. We find evidence for rapid oscillations in six stars: KIC\,6631188, KIC\,7018170, KIC\,10685175, KIC\,11031749,  KIC\,11296437 and KIC\,11409673, and identify each star as chemically peculiar through either pre-existing classifications or spectroscopic measurements. For each star, we identify the principal pulsation mode, and are able to observe several additional pulsation modes in KIC\,7018170. We find that KIC\,7018170 and KIC\,11409673 both oscillate above their theoretical acoustic cutoff frequency, whilst KIC\,11031749 oscillates at the cutoff frequency within uncertainty. All but KIC\,11031749 exhibit strong amplitude modulation consistent with the oblique pulsator model, confirming their mode geometry and periods of rotation.
\end{abstract}




\section{Introduction}
\label{sec:intro}


Since their discovery by \citet{Kurtz197812, Kurtz1982Rapidly}, only 70 rapidly oscillating Ap (roAp) stars have been found \citep{Smalley2015KIC, Joshi2016NainitalCape, Cunha2019Rotation, Balona2019Highfrequencies}. Progress in understanding their pulsation mechanism, abundance, and the origin of their magnetic fields has been hindered by the relatively small number of known roAp stars. A key difficulty in their detection lies in the rapid oscillations themselves, requiring dedicated observations at a short enough cadence to properly sample the oscillations. In this paper, we show that the \kepler\ long-cadence data can be used to detect roAp stars, despite their pulsation frequencies being greater than the Nyquist frequency of the data.


As a class, the chemically peculiar A type (Ap) stars exhibit enhanced features of rare earth elements, such as Sr, Cr and Eu, in their spectra \citep{Morgan1933Evidence}. This enhancement is the result of a stable magnetic field on the order of a few to tens of kG \citep{Mathys2017Ap}, which typically allows for the formation of abundance `spots' on the surface, concentrated at the magnetic poles \citep{Ryabchikova2007Pulsation}. In most, but not all Ap stars, photometric and spectral variability over the rotation cycle can be observed \citep{Abt1995Relation}. Such characteristic spot-based modulation manifests as a low-frequency modulation of the light curve which is readily identified, allowing for the rotation period to be measured \citep[e.g.][]{Drury2017Large}.

The roAp stars are a rare subclass of the Ap stars that exhibit rapid brightness and radial velocity variations with periods between 5 and 24 min and amplitudes up to 0.018 mag in Johnson $B$ \citep{Kurtz2000Introduction, Kochukhov2009Asteroseismology}. They oscillate in high-overtone, low-degree pressure (p) modes \citep{Saio2005Nonadiabatic}. The excitation of high-overtone p-modes, as opposed to the low-overtones of other pulsators in the classical instability strip is suspected to be a consequence of the strong magnetic field -- on the order of a few to tens of kG -- which suppresses the convective envelope at the magnetic poles and increases the efficiency of the opacity mechanism in the region of hydrogen ionisation \citep{Balmforth2001Excitation,Cunha2002Theoretical}. Based on this, a theoretical instability strip for the roAp stars has been published by \citet{Cunha2002Theoretical}. However, discrepancies between the observed and theoretical red and blue edges have been noted, with several roAp stars identified to be cooler than the theoretical red edge. 

A further challenge to theoretical models of pulsations in magnetic stars are oscillations above the so-called acoustic cutoff frequency \citep{Saio2013Pulsation,Holdsworth2018LCO}. In non-magnetic stars, oscillations above this frequency are not expected. However, in roAp stars the strong magnetic field guarantees that part of the wave energy is kept inside the star in every pulsation cycle, for arbitrarily large frequencies \citep{sousaandcunha2008}. For that reason, no theoretical limit exists to the frequency of the modes. Nevertheless, for a mode to be observed, it has to be excited. Models show that the opacity mechanism is capable of exciting modes of frequency close to, but below, the acoustic cutoff frequency. The excitation mechanism for the oscillations above the acoustic cutoff is thought to be turbulent pressure in the envelope regions where convection is no longer suppressed \citep{Cunha2013Testing}.

The magnetic axis of roAp stars is closely aligned with the pulsation axis, with both being inclined to the rotation axis. Observation of this phenomenon led to the development \citep{Kurtz1982Rapidly} and later refinement \citep{Dziembowski1985Frequency,Shibahashi1985Rapid,Shibahashi1985Rotational,Shibahashi1993Theory,Takata1994Selection,Takata1995Effects,Bigot2011Theoretical} of the oblique pulsator model. The roAp stars present a unique testbed for models of magneto-acoustic interactions in stars, and have been widely sought with both ground and space-based photometry.

The launch of the \textit{Kepler} Space Telescope allowed for the detection of oscillations well below the amplitude threshold for ground-based observations, even for stars fainter than 13 magnitude. The vast majority of stars observed by \kepler\ were recorded in long-cadence (LC) mode, with exposures integrated over 29.43\,min. A further allocation of 512 stars at any given time were observed in the short-cadence (SC) mode, with an integration time of 58.85\,s. These two modes correspond to Nyquist limits of 283.21 and 8496.18\,\muhz\, respectively \citep{Borucki2010Kepler}. In its nominal mission, \textit{Kepler} continuously observed around 150 000 stars in LC mode for 4 yr.

The \textit{Kepler} SC data have been used to discover several roAp stars \citep{Kurtz2011First,Balona2011Kepler,Balona2013Unusual,Smalley2015KIC}, and to detect pulsation in previously known roAp stars with the extended K2 mission \citep{Holdsworth2016HD,Holdsworth2018K2}. Until now, only SC data have been used for identification of new roAp stars in the \textit{Kepler} field. However, with the limited availability of SC observation slots, a wide search for rapid oscillators has not been feasible. Although ground-based photometric data have been used to search for roAp stars \citep[e.g.][]{Martinez1991Cape,Joshi2005NainitalCape,Paunzen2012Hvar,Holdsworth2014Highfrequency}, most previous work in using \kepler\ to identify such stars has relied solely on SC observations of targets already known to be chemically peculiar. The number of targets in the \kepler\ field that possess LC data far outweigh those with SC data, but they have been largely ignored in the search for new roAp stars. 

\begin{figure*}
    \centering
    \includegraphics[width=\textwidth]{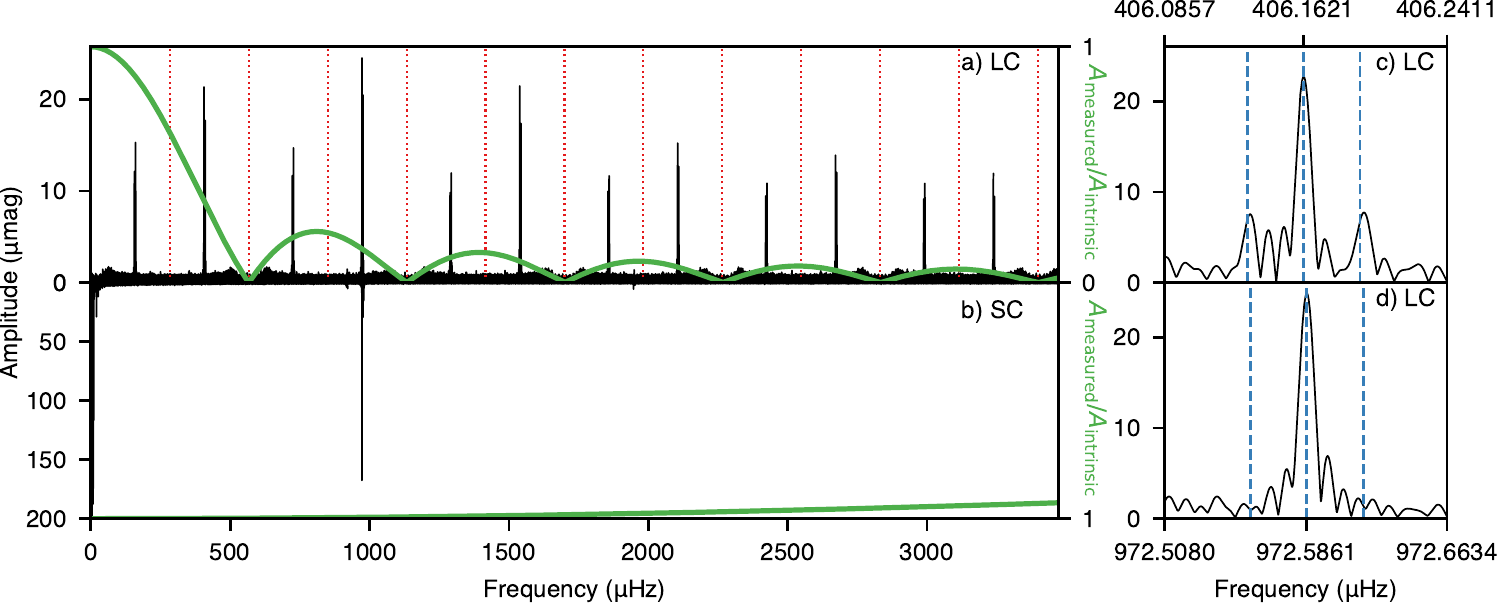}
    \caption{Amplitude spectra of the \textbf{a)} long- and \textbf{b)} short-cadence \kepler\ data of KIC\,10195926, a previously known roAp star \citep{Kurtz2011First}. The primary oscillation is detectable even though it lies far above the Nyquist frequency (shown in integer multiples of the Nyquist frequency as red dotted lines) for the LC data. The green curves show the ratio of measured to intrinsic amplitudes in the data, showing the effects of apodization. \textbf{c)} shows the aliased signal at 406.2\,\muhz\ of the true oscillation at \textbf{d)}, 972.6\,\muhz, distinguishable by both the \kepler\ orbital separation frequency (dashed blue lines) and maximum amplitudes. The Python code P{\sc{yquist}} has been used to plot the apodization \citep{Bell2017Destroying}.}
    \label{fig:10195926}
\end{figure*}

The key difficulty in searching for rapid oscillations in the LC data is that each pulsation frequency in the Fourier spectrum is accompanied by many aliases, reflected around integer multiples of the sampling frequency. Despite this, it has previously been shown by \citet{Murphy2013SuperNyquist} that the Nyquist ambiguity in the LC data can be resolved as a result of the barycentric corrections applied to \kepler\ time stamps, leading to a scenario where Nyquist aliases can be reliably distinguished from their true counterparts even if they are well above or below the nominal Nyquist limit. The barycentric corrections modulate the cadence of the photometric observations, so that all aliases above the Nyquist limit appear as multiplets split by the orbital frequency of the \kepler\ telescope ($1/372.5$\,d, 0.03\,\muhz). Furthermore, the distribution of power in Fourier space ensures that in the absence of errors, the highest peak of a set of aliases will always be the true one. An example of distinguishing aliases is shown in Fig.~\ref{fig:10195926} for the known roAp star KIC\,10195926. The true pulsation is evident as the highest peak in the LC data, and is not split by the \kepler\ orbital frequency.

This technique, known as super-Nyquist asteroseismology, has previously been used with red giants and solar-like oscillators on a case-by-case basis \citep{Chaplin2014SuperNyquist,Yu2016Asteroseismology,Mathur2016Probing}, as well as in combinations of LC data with ground-based observations for compact pulsators \citep{Bell2017Destroying}. Applications in the context of roAp stars have been limited only to frequency verification of SC or other data \citep{Holdsworth2014KIC, Smalley2015KIC}. Our approach makes no assumption about the spectroscopic nature or previous identification of the target, except that its effective temperature lies in the observed range for roAp stars. We further note that super-Nyquist asteroseismology is applicable to the Transiting Exoplanet Survey Satellite \citep[TESS;][]{Ricker2014Transiting} and future space-based missions \citep{Murphy2015Potential, Shibahashi2018SuperNyquist}.

In this paper, we report six new roAp stars whose frequencies are identified solely from their LC data; KIC\,6631188, KIC\,7018170, KIC\,10685175, KIC\,11031749, KIC\,11296437, and KIC\,11409673. These are all found to be chemically peculiar A/F stars with enhanced Sr, Cr, and/or Eu lines.

\section{Observational data \& Analysis}
\subsection{Target selection}

We selected \textit{Kepler} targets with effective temperatures between 6000 and 10\,000\,K according to the `input' temperatures of \citet{Mathur2017Revised}. We significantly extended the cooler edge of our search since few roAp stars are known to lie close to the red edge of the instability strip. We used the \textit{Kepler} LC light curves from Quarters 0 through 17, processed with the PDCSAP pipeline \citep{Stumpe2014Multiscale}, yielding a total sample of 69\,347 stars. We applied a custom-written pipeline to all of these stars that have LC photometry for at least four quarters. Nyquist aliases in stars with time-bases shorter than a full Kepler orbital period (4 quarters) have poorly defined multiplets and were discarded from the sample at run-time \citep[see][for details]{Murphy2013SuperNyquist}.

In addition to the automated search, we manually inspected the light curves of the 53 known magnetic chemically peculiar (mCP) stars from the list of \citet{Hummerich2018Kepler}. These stars have pre-existing spectral classification, requiring only a super-Nyquist oscillation to be identified as a roAp star.

\subsection{Pipeline}

The pipeline was designed to identify all oscillations between 580 and 3500\,\muhz\ in the \kepler\ LC data, by first applying a high-pass filter to the light curve, removing both the long-period rotational modulation between $0$ and $50$\,\muhz\ and low-frequency instrumental artefacts. The high-pass filter reduced all power in this given range to noise level. The skewness of the amplitude spectrum values, as measured between 0 and 3500\,\muhz, was then used as a diagnostic for separating pulsators and non-pulsators, following \citet{Murphy2019Gaiaderived}. Stars with no detectable pulsations, either aliased or otherwise, tend to have a skewness lower than unity, and were removed from the sample.

After filtering out non-pulsators, each frequency above 700\,\muhz\ at a signal-to-noise ratio (SNR) greater than 5 was then checked automatically for sidelobes. These sidelobes are caused by the uneven sampling of \textit{Kepler}'s data points once barycentric corrections to the time stamps have been made, as seen in Fig.~\ref{fig:10195926}. A simple peak-finding algorithm was used to determine the frequency of highest amplitude for each frequency in the set of aliases, which was further refined using a three-point parabolic interpolation to mitigate any potential frequency drift. The frequencies above a SNR of 5 were then deemed aliases if their sidelobes were separated by the \kepler\ orbital frequency for a tolerance of $\pm$0.002\,\muhz. Frequencies that did not display evidence of Nyquist aliasing were then flagged for manual inspection. 

The general process for the pipeline can be summarised as follows:
\begin{enumerate}
    \item High-pass filter the light curve and calculate the skewness of the amplitude spectrum between 0 and 3500\,\muhz; if skewness is less than unity, move to next star.
    \item For each peak greater than 700\,\muhz\ with a SNR above 5, identify all sidelobes and determine whether they are separated by the \kepler\ orbital frequency.
   \item If at least one peak is not an alias, flag the star for manual inspection.
\end{enumerate}


\subsection{Apodization}

The high-pass filter was designed to remove all signals between 0 and 50\,\muhz. This had the additional effect of removing the reflected signals at integer multiples of the sampling frequency, regardless of whether they are aliased or genuine. As a result, the pipeline presented here cannot reliably identify oscillations close to integer multiples of the sampling frequency (2$\nu_{\rm Nyq}$). 

However, we note that if any of these stars are indeed oscillating in these regions, or even above the Nyquist frequency, the measured amplitude will be highly diminished as a result of the non-zero duration of \kepler\ integration times, a phenomenon referred to as apodization \citep{Murphy2015Investigating,Hekker2017Giant} or phase smearing \citep{Bell2017Destroying}. The amplitudes measured from the data ($A_{\rm measured}$) are smaller than their intrinsic amplitudes in the \kepler\ filter by a factor of $\eta$,
\begin{eqnarray}
    \eta = \dfrac{A_{\rm measured}}{A_{\rm intrinsic}} = {\rm sinc}\Big[\dfrac{\nu}{2\nu_{\rm Nyq}} \Big],
\end{eqnarray}
where $\nu$ and $\nu_{\rm Nyq}$ are the observed and Nyquist frequencies, respectively. This equation shows that frequencies lying near integer multiples of the sampling frequency are almost undetectable in \textit{Kepler} and other photometric campaigns. The factor $\eta$ is shown as the green curves in Fig.~\ref{fig:10195926}. For the results in Sec.~\ref{sec:results}, both measured and intrinsic amplitudes are provided.

\section{Results}
\label{sec:results}

Each star found to have non-alias high-frequency pulsations by the pipeline was manually inspected. Of the flagged candidates, 4 were previously identified roAp stars in the \kepler\ field, KIC\,10195926 \citep{Kurtz2011First}, KIC\,10483436 \citep{Balona2011Rotation}, KIC\,7582608 \citep{Holdsworth2014KIC}, and KIC\,4768731 \citep{Smalley2015KIC}. The fifth previously known \kepler\ roAp star KIC\,8677585 \citep{Balona2013Unusual}, was not identified by the pipeline, due to the primary frequency of 1659.79\,\muhz\ falling just within range of the filtered region. We further identified one more high-frequency oscillator during manual inspection of the 53 stars in the mCP sample of \citet{Hummerich2018Kepler}.

For all six newly identified stars, we calculated an amplitude spectrum in the frequency range around the detected pulsation, following the method of \citet{Kurtz1985Algorithm}. The frequencies were then optimised by non-linear least-squares. The signal-to-noise ratio (SNR) of the spectrum was calculated for the entire light curve by means of a box kernel convolution of frequency width 23.15\,\muhz\ (2\,\cd), as implemented in the L{\sc{ightkurve}} Python package \citep{lightkurve}.

\subsection{Stellar properties}

The properties of the six new high-frequency oscillators examined in this work are provided in Table~\ref{tab:sample}. Temperatures were obtained from LAMOST DR4 spectroscopy \citep{Zhao2012LAMOST}. Since the temperatures of roAp stars are inherently difficult to measure as a result of their anomalous elemental distributions \citep{Matthews1999Parallaxes}, we inflated the low uncertainties in the LAMOST catalogue ($\sim$40\,K) to a fixed 300\,K. For the one star with an unusable spectrum in LAMOST, KIC\,6631188, we took the temperature from the stellar properties catalogue of \cite{Mathur2017Revised}. 
We derived apparent magnitudes in the SDSS $g-$band by re-calibrating the KIC $g-$ and $r-$bands following equation~1 of \citet{Pinsonneault2012Revised}. Distances were obtained from the Gaia DR2 parallaxes using the normalised posterior distribution and adopted length scale model of \citet{Bailer-Jones2018Estimating}. This produced a distribution of distances for each star, from which Monte Carlo draws could be sampled. Unlike \citet{Bailer-Jones2018Estimating}, no parallax zero-point correction has been applied to our sample, since it has previously been shown by \citet{Murphy2019Gaiaderived} to not be appropriate for \textit{Kepler} A stars.

\begin{table}
	\centering
	\caption{Properties of the 6 new roAp stars.}
	\begin{tabular}{lccccr} 
		\hline
		KIC & $g$ Mag  & T$_{\rm eff}$ (K) & $\log{\rm L/\rm L_\odot}$ & $\rm M/\rm M_\odot$ \\
		\hline
		6631188     & 13.835 & 7700\,$\pm$\,300 & 1.124\,$\pm$\,0.034 & 1.83\,$\pm$\,0.25 \\
		7018170     & 13.335 & 7000\,$\pm$\,300 & 0.987\,$\pm$\,0.026 & 1.69\,$\pm$\,0.25 \\
		10685175    & 12.011& 8000\,$\pm$\,300   & 0.896\,$\pm$\,0.022  & 1.65\,$\pm$\,0.25 \\
		11031749    & 12.949 & 7000\,$\pm$\,300 & 1.132\,$\pm$\,0.041 & 1.78\,$\pm$\,0.25 \\
		11296437    & 11.822 & 7000\,$\pm$\,300 & 1.055\,$\pm$\,0.018 & 1.73\,$\pm$\,0.25 \\
		11409673    & 12.837 & 7500\,$\pm$\,300 & 1.056\,$\pm$\,0.031 & 1.75\,$\pm$\,0.25 \\
		\hline
	\end{tabular}
	\label{tab:sample}
\end{table}

Standard treatment of bolometric corrections \citep[e.g.][]{Torres2010Use} are unreliable for Ap stars, due to their anomalous flux distributions. Working in SDSS $g$ minimises the bolometric correction, since the wavelength range is close to the peak of the spectral energy distribution of Ap stars. We obtained $g$-band bolometric corrections using the {\sc IsoClassify} package \citep{Huber2017Asteroseismology}, which interpolates over the {\sc mesa} Isochrones \& Stellar Tracks (MIST) tables \citep{Dotter2016MESA} using stellar metallicities, effective temperatures, and surface gravities obtained from \citet{Mathur2017Revised}. Extinction corrections of \citet{Green2018Galactic} as queried through the {\sc Dustmaps} python package \citep{Green2018Dustmaps}, were applied to the sample. The corrections were re-scaled to SDSS $g$ following table~A1 of \citet{Sanders2018Isochronea}. To calculate luminosities, we followed the methodology of \citet{Murphy2019Gaiaderived}, using a Monte Carlo simulation to obtain uncertainties. Masses were obtained via an interpolation over stellar tracks, and are discussed in more detail in Sec.~\ref{sec:modelling}.

\section{Discussion of individual stars}
\label{sec:pulsations}
\subsection{KIC~6631188}

\begin{figure*}
    \centering
    \includegraphics[width=\linewidth]{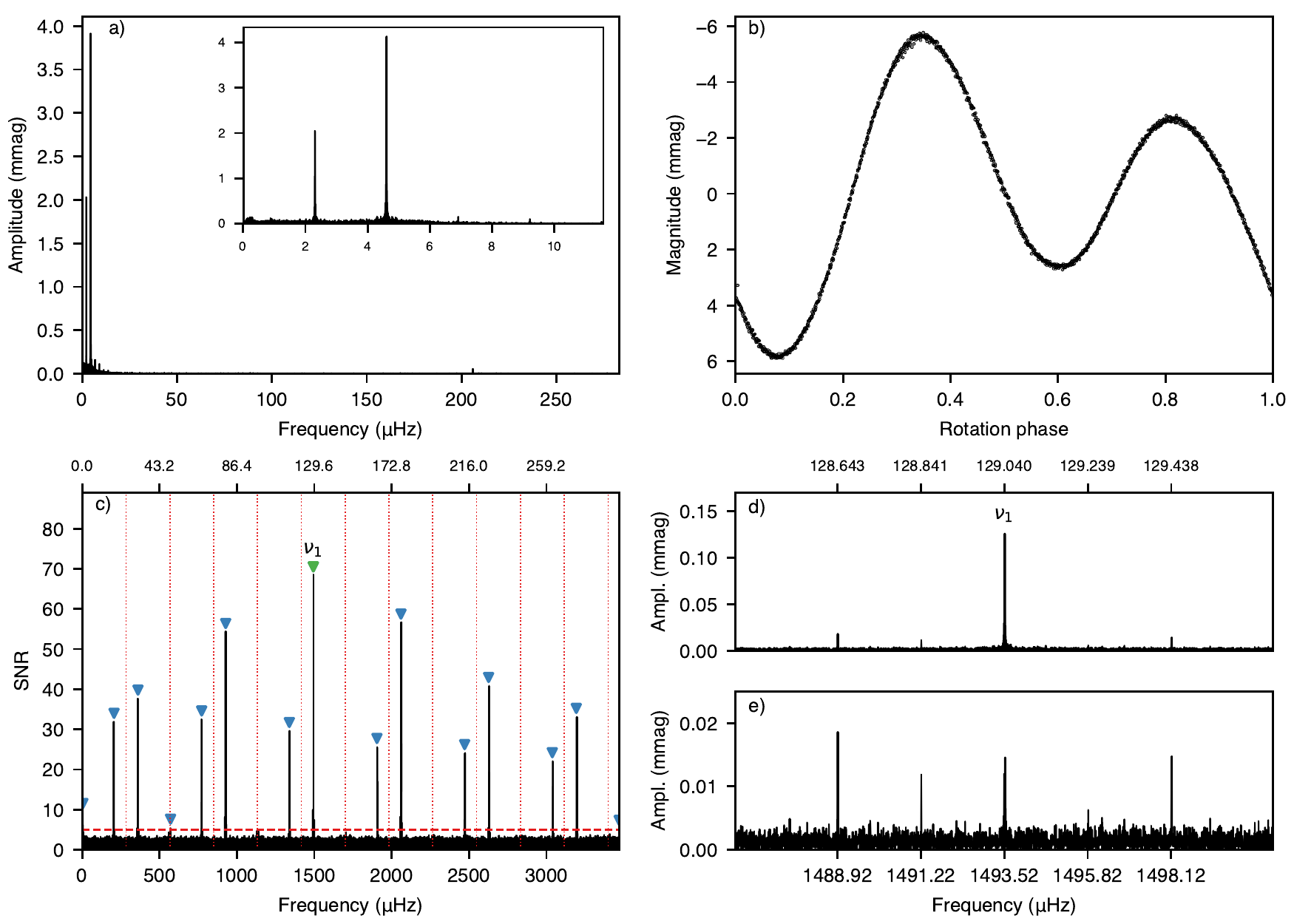}
    \caption{\textbf{a)} Amplitude spectrum of KIC\,6631188 out to the Nyquist frequency of 283.2\,\muhz. The inset shows the low-frequency region with peaks due to rotation. \textbf{b)} Light curve folded at the rotation period of 5.03 d and binned at a factor of 50:1. \textbf{c)} Amplitude spectrum of KIC\,6631188 after a high-pass filter has removed the low-frequency signals -- the true oscillation frequency of 1493.52\,\muhz\ has the highest amplitude (green). All other peaks flagged as aliases above the 5 SNR are marked in blue. The red dashed lines denote integer multiples of the Nyquist frequency. \textbf{d)} Zoomed region of the primary frequency before, and \textbf{e)}, after pre-whitening $\nu_1$. The residual power in $\nu_1$ is due to frequency variability. The four sidelobes are due to rotational modulation of the pulsation amplitude (see text). The top x-axis, where shown, is the corresponding frequency in d$^{-1}$.}
    \label{fig:6631188}
\end{figure*}

\begin{table*}
    \centering
    \caption{Non-linear least squares fits of the rotation frequency and the pulsation multiplets. The zero-points of the fits were chosen to be the centre of each light curve. Rotational frequencies, \nurot, are calculated from the low-frequency portion of the unfiltered light curve when available, whereas the oscillation frequencies are calculated on the high-pass filtered light curve. $\delta \nu$ is the difference in frequency between the current and previous row. $^\dagger$Rotation has been calculated from amplitude modulation of its frequencies (cf. Sec.~\ref{sec:amp_variability}).}
    \begin{tabular}{
        l 
        l 
        r@{$\,\pm\,$}l 
        r@{$\,\pm\,$}l 
        r@{$\,\pm\,$}l 
        r@{$\,\pm\,$}l 
        c 
        r 
        }
        \hline
        KIC & Label & \multicolumn{2}{c}{Frequency} & \multicolumn{2}{c}{Amplitude$_{\rm \,\,measured}$}    & \multicolumn{2}{c}{Amplitude$_{\rm \,\,intrinsic}$}    & \multicolumn{2}{c}{Phase} & $\delta \nu$ & $\delta \nu / \nu_{\rm rot}$\\
            &       & \multicolumn{2}{c}{(\muhz)}   & \multicolumn{2}{c}{(mmag)}                          & \multicolumn{2}{c}{(mmag)}                      &\multicolumn{2}{c}{(rad)} & (\muhz) \\
        \hline
        6631188 & \nurot    & 2.300\,47&0.000\,04   & 2.053&0.016 &   2.053&0.016
        \vspace{0.05cm} \\
        &$\nu_1$ - 2\nurot  &  1488.918\,89&0.000\,25 &  0.018&0.001   &   0.163&0.009       & 2.171&0.058    &\\
        &$\nu_1$ - \nurot   &  1491.219\,21&0.000\,42 &  0.011&0.001   &   0.100&0.009    & -2.601&0.095   & 2.300&1.000\\
        &$\nu_1$            &  1493.519\,47&0.000\,04 &  0.123&0.001  &   1.119&0.009   & 0.672&0.009    &2.300&1.000\\
        &$\nu_1$ + \nurot   &  1495.819\,89&0.000\,74 &  0.006&0.001    &   0.058&0.009   & 0.906&0.169    &2.300&1.000\\
        &$\nu_1$ + 2\nurot  &  1498.121\,02&0.000\,33 &  0.014&0.001   &   0.130&0.009   & -2.092&0.074   &2.301&1.000\\
        \hline
        7018170 & \nurot    & 0.1591&0.0054{$^\dagger$} 
        \vspace{0.05cm} \\
        &$\nu_1$ - 2\nurot   & 1944.982\,10&0.001\,23 &  0.008&0.001    &  0.092&0.013 &     3.128&0.322 &   \\
        &$\nu_1$ - \nurot    & 1945.142\,81&0.000\,67 &  0.015&0.001    &  0.170&0.013 &     -3.038&0.176 &   0.161&1.010\\
        &$\nu_1$             & 1945.301\,73&0.000\,22 &  0.047&0.001    &  0.513&0.013 &     -2.687&0.058 &   0.159&0.999\\
        &$\nu_1$ + \nurot    & 1945.454\,78&0.003\,06 &  0.003&0.001    &  0.037&0.013 &     -4.674&0.806 &   0.153&0.962\\
        &$\nu_1$ + 2\nurot   & 1945.621\,68&0.003\,33 &  0.003&0.001    &  0.034&0.013 &     -1.768&0.874 &   0.167&1.049\\
        \vspace{0.05cm} \\
        &$\nu_2$ - \nurot    & 1920.120\,72&0.001\,41 &  0.007&0.001    &   0.082&0.013 &    0.738&0.369 &  \\
        &$\nu_2$             & 1920.278\,31&0.000\,95 &  0.011&0.001    &   0.123&0.013 &    0.504&0.249 &  0.158&0.991\\
        &$\nu_2$ + \nurot    & 1920.439\,44&0.002\,47 &  0.004&0.001    &   0.047&0.013 &    0.268&0.648 &  0.161&1.013\\
        \vspace{0.05cm} \\
        &$\nu_3$ - \nurot    & 1970.165\,86&0.001\,69 &  0.006&0.001    &  0.066&0.013 &     -1.657&0.444 &   \\
        &$\nu_3$             & 1970.324\,09&0.001\,47 &  0.007&0.001    &  0.077&0.013 &     -2.795&0.385 &   0.158&0.995\\
        &$\nu_3$ + \nurot    & 1970.483\,30&0.001\,83 &  0.006&0.001    &  0.061&0.013 &     -2.552&0.481 &   0.159&1.001\\
        \hline
        10685175  & \nurot  & 3.731\,18&0.000\,01     &   4.951&0.010 & 4.951&0.010
        \vspace{0.05cm} \\
        &$\nu_1$ - 2\nurot  &  2775.545\,47&0.003\,38 &  0.004&0.003   &   0.191&0.150       & 1.077&0.800    &\\
        &$\nu_1$ - \nurot   &  2779.225\,90&0.002\,03 &  0.006&0.003   &   0.337&0.161    & -1.803&0.481   & 3.680&0.986\\
        &$\nu_1$            &  2783.008\,00&0.000\,96 &  0.013&0.003  &   0.765&0.173   & 2.953&0.228    &3.782&1.014\\
        &$\nu_1$ + \nurot   &  2786.689\,62&0.003\,00 &  0.004&0.003    &   0.266&0.187   & 0.437&0.708   &3.682&0.987\\
        &$\nu_1$ + 2\nurot  &  2790.983\,95&0.002\,84 &  0.005&0.003   &   0.310&0.206  & -2.863&0.671   &4.294&1.151\\
        \hline
        11031749 & $\nu_1$ & 1372.717\,24&0.000\,16 & 0.0261&0.0006 &0.205&0.005    & 0.899 &0.021\\
        \hline
        11296437 & \nurot  & 1.624\,58&0.000\,01 & 1.705&0.002 & 1.705&0.002
        \vspace{0.05cm} \\
        &$\nu_1$ - \nurot      & 1408.152\,04&0.000\,44 &  0.0026&0.0003    &  0.020&0.002  & -2.229&0.100 \\
        &$\nu_1$               & 1409.776\,71&0.000\,02 &  0.0450&0.0003   &  0.352&0.002  & 2.437&0.006    &1.625&1.000\\
        &$\nu_1$ + \nurot      & 1411.402\,13&0.000\,49 &  0.0023&0.0003    &  0.018&0.002  & 0.823&0.112   &1.625&1.001\\
        \vspace{0.05cm} \\
        &$\nu_2$               & 126.791\,38&0.000\,05  & 0.0222&0.0003    &  0.0242&0.0003 & -1.716&0.012 \\
        &$\nu_3$               & 129.151\,22&0.000\,04  &  0.0317&0.0003    &  0.0345&0.0003 & 2.080&0.008 \\
        \hline
        11409673 & \nurot & 0.940\,16&0.000\,02 & 0.865&0.006 & 0.865&0.006
        \vspace{0.05cm} \\
        &$\nu_1$ - \nurot       & 2499.985\,30&0.001\,01 & 0.021&0.001 & 0.307&0.018    &-1.571&0.057& \\
        &$\nu_1$                & 2500.926\,65&0.003\,21  & 0.007&0.001 & 0.097&0.018     &0.667&0.199  &0.941&1.001 \\
        &$\nu_1$ + \nurot       & 2501.866\,33&0.001\,11 & 0.019&0.001 & 0.280&0.018    &-1.757&0.069   &0.940&0.999 \\
        \hline
    \end{tabular}
    \label{tab:pulsators}
\end{table*}
KIC\,6631188 has previously been identified as a rotational variable with a period of 5.029\,d \citep{Reinhold2013Rotation}, or 2.514\,d \citep{Reinhold2015Rotation}. The unfiltered light curve of KIC\,6631188 shows a series of low-frequency harmonic signals beginning at multiples of 2.30\,\muhz\ (Fig.~\ref{fig:6631188}). Although the highest amplitude signal corresponds to a rotational period of 2.514\,d, the true rotation period was confirmed by folding the light curve on the 2.30\,\muhz\ frequency, yielding a period of 5.03117\,$\pm$\,0.00004\,d. The folded light curve shows clear double-wave spot-based modulation, implying that both magnetic poles are observed.

After high-pass filtering the light curve, the primary pulsation frequency of 1493.52\,\muhz\ is observable in the super-Nyquist regime. We see evidence for rotational splitting through the detection of a quintuplet, indicating an $\ell=2$ or distorted $\ell=1$ mode. It seems likely that the star is a pure quadrupole pulsator, unless an $\ell=1$ mode is hidden at an integer multiple of the sampling frequency -- where its amplitude would be highly diminished as a result of apodization. It is also possible that other modes are of low intrinsic amplitude, making their detection in the super-Nyquist regime difficult. We can measure the rotational period of KIC\,6631188 from the sidelobe splitting as 5.0312\,$\pm$\,0.0003\,d in good agreement with the low-frequency signal. We list the pulsation and rotational frequencies in Table~\ref{tab:pulsators}.

We are able to provide further constraints on the geometry of the star by assuming that the rotational sidelobes are split from the central peak by exactly the rotation frequency of the star. We chose a zero-point in time such that the phases of the sidelobes were equal, and then applied a linear least squares fit to the data. For a pure non-distorted mode, we expect the phases of all peaks in the multiplet to be the same. We find that the phases are not identical, implying moderate distortion of the mode (Table~\ref{tab:6631188_forcefit}).

\begin{table}
    \centering
    \caption{Linear least squares fit to the pulsation and force-fitted sidelobes in KIC\,6631188. The zero-point for the fit is BJD 2455692.84871, and has been chosen as such to force the first pair of sidelobe phases to be equal.}
    \label{tab:6631188_forcefit}
    \begin{tabular}{
    lccc
    }
        \hline
        ID  & {Frequency}   & {Amplitude$_{\rm \,\,intrinsic}$}  & {Phase}  \\
            & {(\muhz)}     & {(mmag)}                    & {(rad)}\\
        \hline
        $\nu_1-2$\nurot & 1488.9185 & 0.163\,$\pm$\,0.009   & 1.579\,$\pm$\,0.058\\
        $\nu_1-$\nurot  & 1491.2190 & 0.100\,$\pm$\,0.009    & 1.337\,$\pm$\,0.095\\
        $\nu_1$         & 1493.5195 & 1.121\,$\pm$\,0.009  & 2.845\,$\pm$\,0.009\\
        $\nu_1+$\nurot  & 1495.8199 & 0.058\,$\pm$\,0.009    & 1.337\,$\pm$\,0.167\\
        $\nu_1+2$\nurot & 1498.1204 & 0.130\,$\pm$\,0.009   & 2.859\,$\pm$\,0.075\\
        \hline
    \end{tabular}
\end{table}

The oblique pulsator model can also be applied to obtain geometric constraints on the star's magnetic obliquity and inclination angles, $\beta$ and $i$, respectively. The frequency quintuplet strongly suggests that the pulsation in KIC\,6631188 is a quadrupole mode. We therefore consider the axisymmetric quadrupole case, where $\ell=2$ and $m=0$ and apply the relation of \citet{Kurtz1990Rapidly} for a non-distorted oblique quadrupole pulsation in the absence of limb-darkening and spots:
\begin{eqnarray}
    \tan{i}\tan{\beta}= 4 \dfrac{A_{+2}^{(2)}+A_{-2}^{(2)}}{A_{+1}^{(2)}+A_{-1}^{(2)}}.
    \label{eqn:quint}
\end{eqnarray}

\begin{figure}
    \centering
    \includegraphics[width=\linewidth]{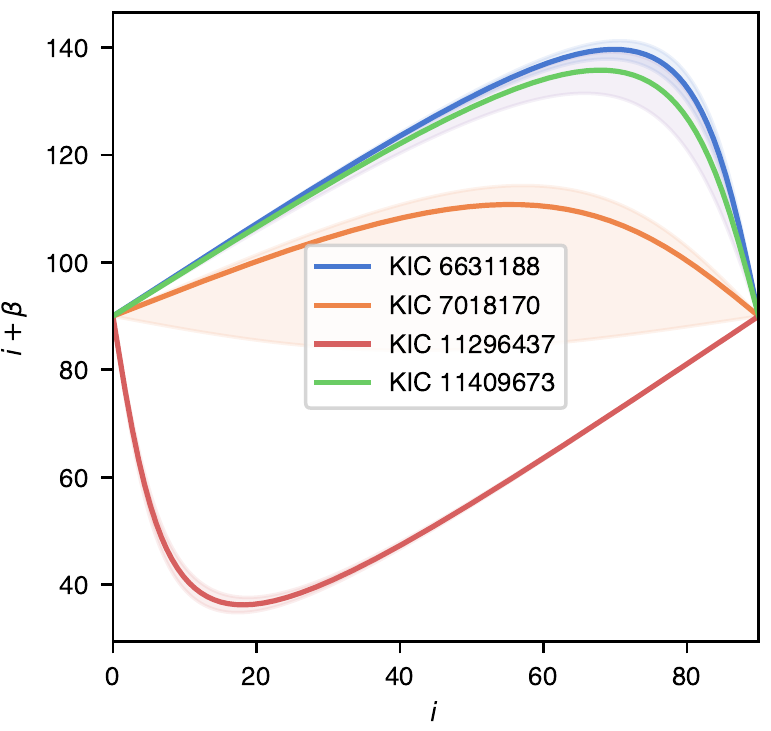}
    \caption{Possible $i + \beta$ combinations for the roAp stars where analysis of the multiplets allows us to set constraints on their geometry. The shaded region marks the uncertainty. In stars for which $i + \beta > 90 ^\circ$, both magnetic poles are observed. For KIC\,7018170, only the primary $\nu_1$ solution has been shown. KIC\,10685175 has been omitted as its uncertainty dominates the figure.}
    \label{fig:ibeta_combo}
\end{figure}

Here $i$ is the rotational inclination angle, $\beta$ is the angle of obliquity between the rotation and magnetic axes, and $A_{\pm1,2}^{(1,2)}$ are the amplitudes of the first and second sidelobes of the quadrupole pulsation. Using the values of Table~\ref{tab:6631188_forcefit}, we find that $\tan{i}\tan{\beta}=7.4\,\pm\,0.7$, and provide a summary of values satisfying this relation in Fig.~\ref{fig:ibeta_combo}. Since $i+\beta \geq 90^\circ$, both pulsation poles should be visible in the light curve over the rotation cycle of the star, a result consistent with observations of the double-wave light curve with spots at the the magnetic poles.

\subsection{KIC~7018170}

\begin{figure*}
    \centering
    \includegraphics[width=\linewidth]{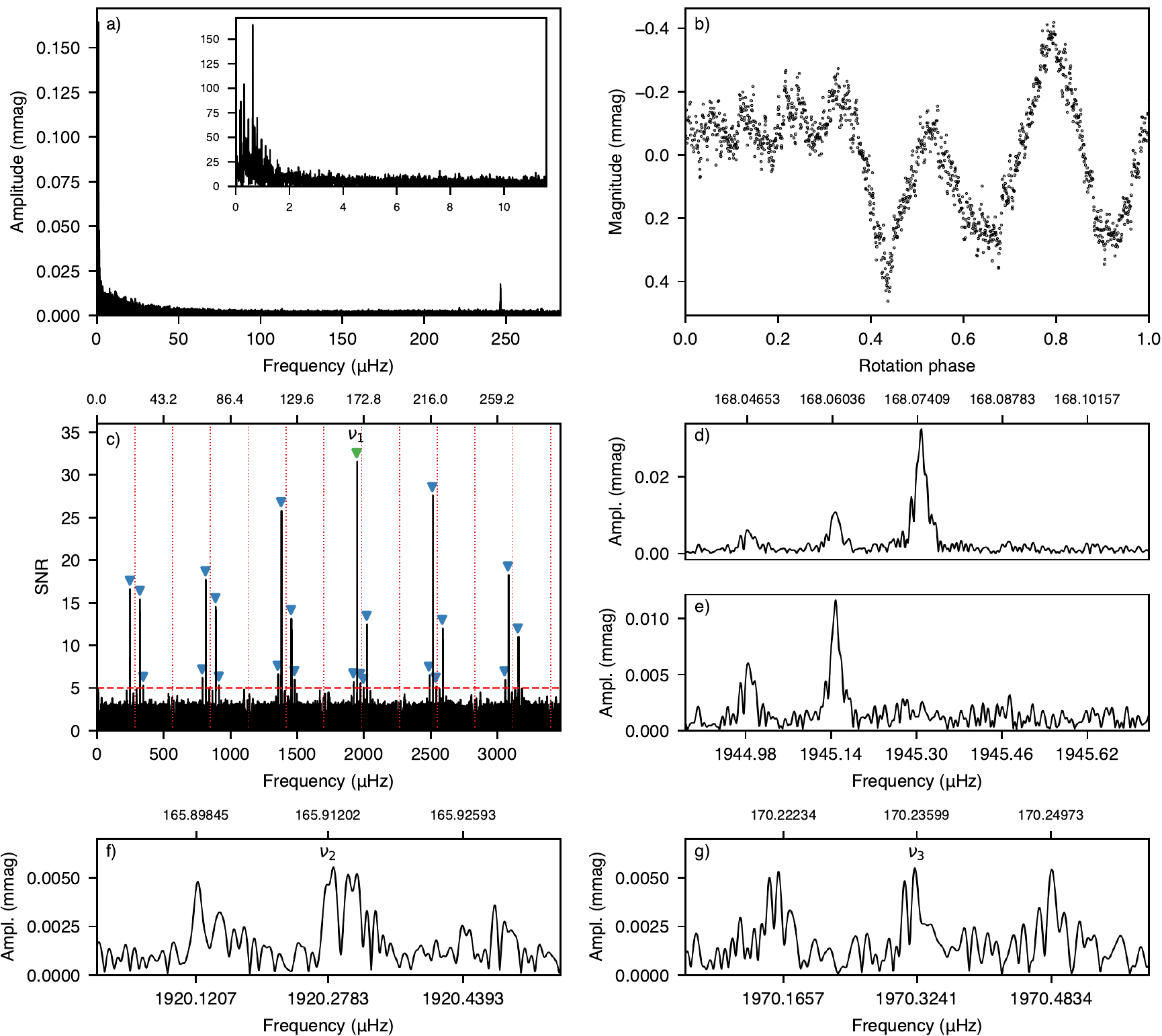}
    \caption{Same as in Fig.~\ref{fig:6631188} for KIC\,7018170. In \textbf{a)} however, the long-period rotational modulation has been largely removed by the PDCSAP flux pipeline leading to a jagged light curve (\textbf{b}). Panels \textbf{f)} and \textbf{g)} show the secondary frequencies $\nu_2$ and $\nu_3$ extracted after manual inspection of the filtered light curve.}
    \label{fig:7018170}
\end{figure*}

The low-frequency variability of KIC\,7018170 exhibits no sign of rotational modulation, which is probably a result of the \mbox{PDCSAP} pipeline removing the long-period variability (Fig.~\ref{fig:7018170}). It is therefore unsurprising that KIC\,7018170 has not been detected as an Ap star in the \kepler\ data -- the automatic removal of low-frequency modulation causes it to appear as an ordinary non-peculiar star in the LC photometry.

The rotational signal is clearly present in the sidelobe splitting of the primary and secondary pulsation frequencies. The high-pass filtered light curve reveals the primary signal, $\nu_1$, at 1945.30\,\muhz, with inspection of the amplitude spectrum revealing two more modes; $\nu_2$ and $\nu_3$, at frequencies of 1920.28 and 1970.32\,\muhz, respectively. All three of these modes are split by 0.16\,\muhz\ which we interpret as the rotational frequency. KIC\,7018170 exhibits significant frequency variability during the second half of the data, which destroys the clean peaks of the triplets. To analyse them in detail, we analysed only the first half of the data where frequency variability is minimal. This provided a good balance between frequency resolution and variability in the data.

To estimate the large frequency separation, $\Delta \nu$, defined as the difference in frequency of modes of the same degree and consecutive radial order, we apply the general asteroseismic scaling relation,
\begin{eqnarray}
    \dfrac{\Delta \nu}{\Delta\nu_{{\odot}}} = \sqrt{\dfrac{\rho}{\rho_{\odot}}} = \dfrac{(M/{\rm M}_{\odot})^{0.5} (T_{\rm eff} / {\rm T}_{\rm eff, \odot})^3}{(L/{\rm L}_{\odot})^{0.75}}
    \label{eqn:dnu}
\end{eqnarray}
with adopted solar values $\Delta \nu_\odot$ = 134.88\,$\pm$\,0.04\,\muhz, and T$_{\rm eff, \odot} = 5777$\,K \citep{Huber2011Testing}. Using the stellar properties in Table~\ref{tab:sample}, we estimate the large separation as 55.19\,$\pm$\,7.27\,\muhz.  The separation from the primary frequency $\nu_1$ to $\nu_2$ and $\nu_3$ are 24.86 and 25.18\,\muhz, respectively, indicating that the observed modes are likely of alternating even and odd degrees. While we can not determine the degrees of the modes from the LC data alone, it suggests that the primary frequency $\nu_1$ is actually a quintuplet with unobserved positive sidelobes.

If $\nu_1$ is instead a triplet, it would have highly asymmetric rotational sidelobe peak amplitudes which the other modes do not exhibit. Asymmetric sidelobe amplitudes are a signature of the Coriolis effect \citep{Bigot2003Are}, but whether this is the sole explanation for such unequal distribution of power in the amplitude spectra can not be known without follow-up observations. $\nu_1$ is thus more likely to be a quintuplet as the large separation would then be 50.04\,\muhz, a value much closer to the expected result. The positive suspected sidelobes at frequencies of 1945.46 and 1945.62\,\muhz\ have a SNR of 1.16 and 1.78 respectively, well below the minimum level required for confirmation. To this end, we provide a fit to the full quintuplet in Table~\ref{tab:pulsators}, but note that the frequencies should be treated with caution.

Assuming that $\nu_2$ and $\nu_3$ are triplets, while $\nu_1$ is a quintuplet, we forced the rotational sidelobes to be equally separated from the pulsation mode frequency by the rotation frequency. Lacking the rotational signal from the low-frequency amplitude spectrum, we instead obtained the rotational frequency by examining the variability in amplitudes of the modes themselves. As the star rotates, the observed amplitudes of the oscillations will modulate in phase with the rotation. We provide a full discussion of this phenomenon in Sec.~\ref{sec:amp_variability}. We obtained a rotation frequency of 0.160\,$\pm$\,0.005\,\muhz, corresponding to a period of 72.7\,$\pm$\,2.5\,d. We used this amplitude modulation frequency to fit the multiplets by linear least squares to test the oblique pulsator model. By choosing the zero-point in time such that the phases of the $\pm\nu_{\rm rot}$ sidelobes of $\nu_1$ are equal, we found that $\nu_1$, does not appear to be distorted, and $\nu_3$ only slightly. $\nu_2$ is heavily distorted, as shown by the unequal phases of the multiplet in Table~\ref{tab:7018170_forcefit}.

\begin{table}
    \centering
    \caption{Linear least squares fit to the pulsation and force-fitted sidelobes in KIC\,7018170. The zero-point for the fit is BJD 2455755.69582, and has been chosen as such to force the sidelobe phases of $\nu_1$ to be equal.}
    \label{tab:7018170_forcefit}
    \begin{tabular}{lccc}
        \hline
        ID    & Frequency & Amplitude$_{\rm \,\,intrinsic}$ & Phase  \\
         & (\muhz) & (mmag) & (rad) \\
        \hline
        $\nu_1$ - 2\nurot   & $1944.98350$ & $0.090\,\pm\,0.013$    & $-2.799\,\pm\,0.141$    \\
        $\nu_1$ - \nurot    & $1945.14259$  & $0.172\,\pm\,0.013$    & $-3.080\,\pm\,0.074$       \\
        $\nu_1$             & $1945.30169$ & $0.510\,\pm\,0.013$    & $-2.683\,\pm\,0.025$         \\
        $\nu_1$ + \nurot    & $1945.46079$ & $0.031\,\pm\,0.013$    & $-3.080\,\pm\,0.403$         \\
        $\nu_1$ + 2\nurot   & $1945.61988$ & $0.033\,\pm\,0.013$    & $-2.170\,\pm\,0.389$         \\
        $\nu_2$ - \nurot    & $1920.11921$ & $0.081\,\pm\,0.013$    & $\phantom{-}0.407\,\pm\,0.161$    \\
        $\nu_2$             & $1920.27831$  & $0.122\,\pm\,0.013$    & $\phantom{-}0.522\,\pm\,0.107$       \\
        $\nu_2$ + \nurot    & $1920.43741$ & $0.046\,\pm\,0.013$    & $-0.109\,\pm\,0.283$         \\
        $\nu_3$ - \nurot    & $1970.16500$ & $0.066\,\pm\,0.013$    & $-1.828\,\pm\,0.192$    \\
        $\nu_3$             & $1970.32410$  & $0.076\,\pm\,0.013$    & $-2.786\,\pm\,0.165$       \\
        $\nu_3$ + \nurot    & $1970.48320$ & $0.061\,\pm\,0.013$    & $-2.562\,\pm\,0.206$\\
        \hline
    \end{tabular}
\end{table}

We can again constrain the inclination and magnetic obliquity angles for the modes. In the case of a pure dipole triplet,
\begin{eqnarray}
    \tan{i}\tan{\beta} = \dfrac{A_{+1}^{(1)}+A_{-1}^{(1)}}{A_{0}^{(1)}},
    \label{eqn:dipole}
\end{eqnarray}
where again $A_{\pm1}^{(1)}$ are the dipole sidelobe amplitudes, and $A_{0}^{(1)}$ is the amplitude of the central peak. Using Table~\ref{tab:7018170_forcefit}, we find that $\tan{i}\tan{\beta} = 1.0\,\pm\,0.2$, and $\tan{i}\tan{\beta} = 1.7\,\pm\,0.4$ for $\nu_2$, and $\nu_3$, respectively. Using Eqn.\,\ref{eqn:quint}, we find $\tan{i}\tan{\beta} = 2.4\,\pm\,0.4$ for $\nu_1$, which agrees with $\nu_3$ within the large errors, while disagreeing with $\nu_2$, which appears to be $\pi$\,rad out of phase. We provide a summary of values satisfying these relations in Fig.~\ref{fig:ibeta_combo}.

\subsection{KIC~10685175}

\begin{figure*}
    \centering
    \includegraphics[width=\linewidth]{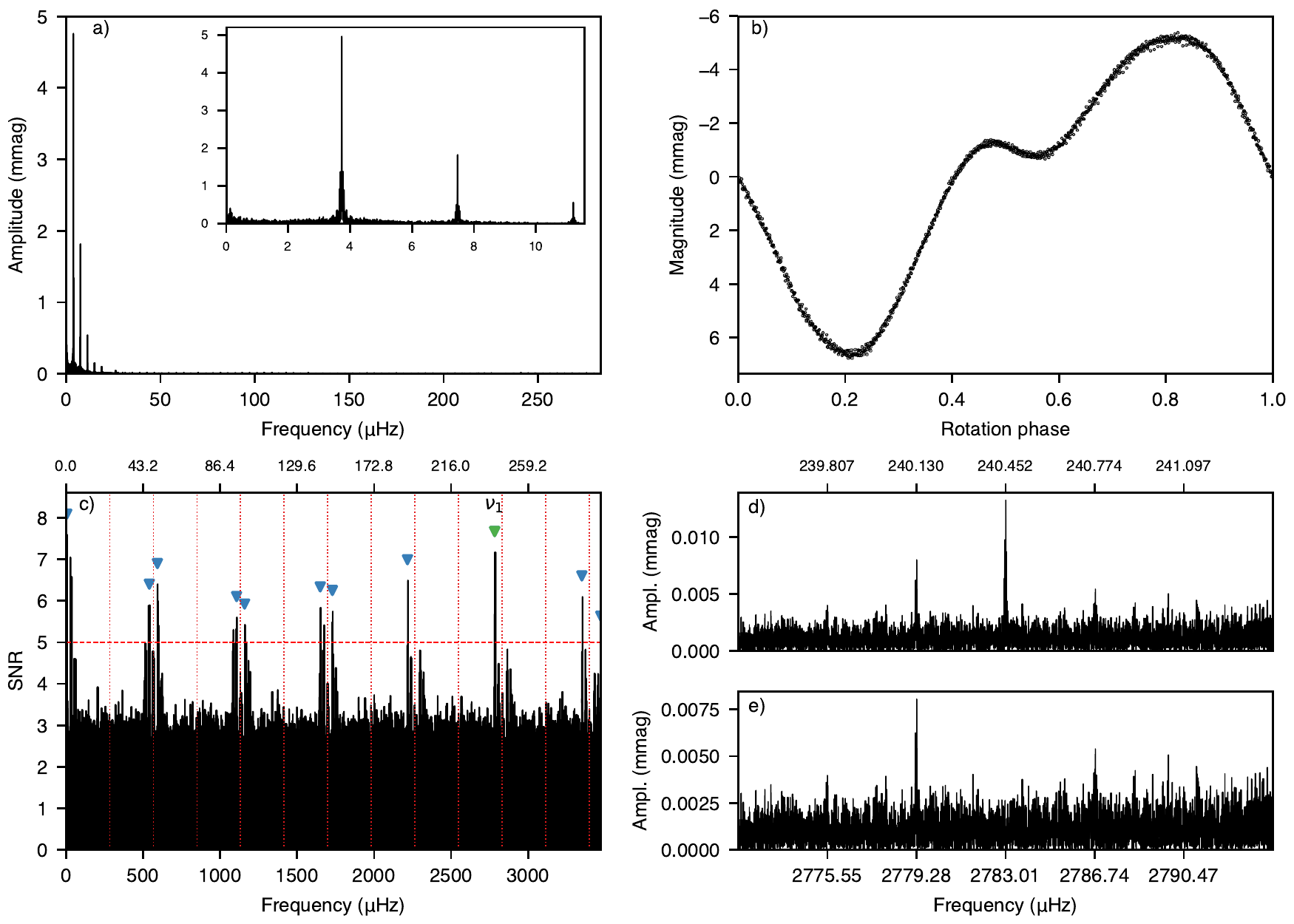}
    \caption{\textbf{a)} Amplitude spectrum of KIC\,10685175 out to the nominal Nyquist frequency. The inset shows the low-frequency region of the spectrum corresponding to the rotation frequency. \textbf{b)} Light curve folded at the rotation period of 5.03 d and binned at a ratio of 50:1. \textbf{c)} Amplitude spectrum of KIC\,10685175 after pre-whitening -- the true oscillation frequency of 2783.00\,\muhz\ can be observed as the signal of maximum amplitude (green). All other peaks flagged as aliases above the 5 SNR are marked in blue. The red dashed lines denote integer multiples of the Nyquist frequency. \textbf{d)} Zoomed region of the primary frequency before, and \textbf{e)}, after pre-whitening $\nu_1$.}
    \label{fig:10685175}
\end{figure*}

KIC\,10685175 was detected by manual inspection of the mCP stars from \citet{Hummerich2018Kepler}, and was not flagged by the pipeline. The star shows obvious rotational modulation in the low-frequency region of the amplitude spectrum (Fig.~\ref{fig:10685175}). The period of rotation, 3.10198\,$\pm$\,0.00001\,d was determined from the low-frequency signal at 0.322\,\muhz.

To study the roAp pulsations, we subtracted the rotational frequency and its first 30 harmonics. Although the amplitude spectrum is too noisy to reveal \kepler\ orbital sidelobe splitting, the true peak is evident as the signal with the highest power: 2783.01\,\muhz. This frequency lies close to a multiple of the \kepler\ sampling rate, and thus has a highly diminished amplitude. The primary frequency appears to be a quintuplet split by the rotational frequency. However, the low SNR of the outermost rotational sidelobes necessitates careful consideration. The outer sidelobes, at frequencies of 2775.55 and 2790.47\,\muhz\ have a SNR of 1.28 and 0.49 respectively. Similar to KIC\,7018170, we provide a fit to the full suspected quintuplet in Table~\ref{tab:pulsators}, but again note that the frequencies of the outermost sidelobes should be treated with caution. If we are to assume that the pulsation is a triplet, while ignoring the outer sidelobes, then Eqn.\,\ref{eqn:dipole} yields a value of $\tan{i}\tan{\beta}=0.9\,\pm\,0.4$, implying that $i+\beta<90^\circ$ and that only one pulsation pole is observed. The large uncertainty however indicates that either one or two poles can be observed if the pulsation is modelled as a triplet. On the other hand, if we consider the star as a quadrupole pulsator, we obtain a value of $\tan{i}\tan{\beta}=1.7\,\pm\,1.6$, a result almost completely dominated by its uncertainty.

We again investigate the distortion of the mode by assuming that the multiplet is split by the rotation frequency, and find that all phases agree within error, implying minimal distortion of the mode (Table~\ref{tab:10685175_forcefit}). However, it should be noted that the low SNR of the spectrum greatly inflates the uncertainties on the amplitudes and phases of the fit.

\begin{table}
    \centering
    \caption{Linear least squares fit to the pulsation and force-fitted sidelobes in KIC\,10685175. The zero-point for the fit is BJD 2455689.78282, and has been chosen as such to force the first pair of sidelobe phases to be equal.}
    \label{tab:10685175_forcefit}
    \begin{tabular}{
    lccc
    }
        \hline
        ID  & {Frequency}   & {Amplitude$_{\rm \,\,intrinsic}$}  & {Phase}  \\
            & {(\muhz)}     & {(mmag)}                    & {(rad)}\\
        \hline
        $\nu_1-2$\nurot & 2775.54564 & 0.192\,$\pm$\,0.150   & 0.842\,$\pm$\,0.782\\
        $\nu_1-$\nurot  & 2779.27682 & 0.432\,$\pm$\,0.161    & 0.324\,$\pm$\,0.373\\
        $\nu_1$         & 2783.00800 & 0.764\,$\pm$\,0.173  & 0.585\,$\pm$\,0.226\\
        $\nu_1+$\nurot  & 2786.73919 & 0.243\,$\pm$\,0.187    & 0.324\,$\pm$\,0.772\\
        $\nu_1+2$\nurot & 2790.47037 & 0.099\,$\pm$\,0.204   & 0.049\,$\pm$\,2.051\\
        \hline
    \end{tabular}
\end{table}

\subsection{KIC~11031749} 

\begin{figure*}
    \centering
    \includegraphics[width=\linewidth]{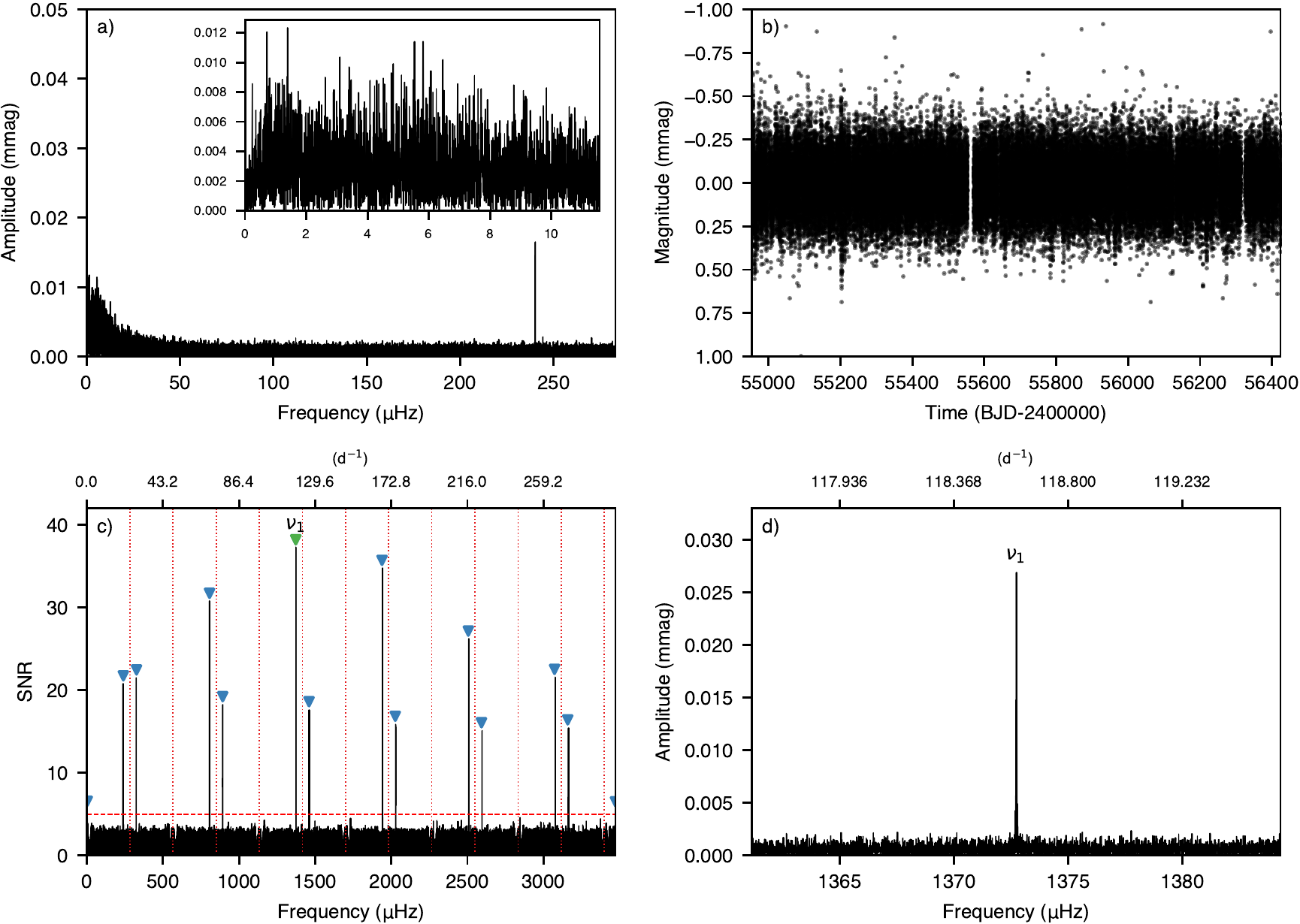}
    \caption{Same as Fig.~\ref{fig:6631188} for KIC\,11031749. No rotation signal can be observed in either \textbf{a)} the low-frequency amplitude spectrum or \textbf{d)} rotational splitting. We provide the full 4 yr light curve in lieu of a folded light curve (\textbf{b}).}
    \label{fig:11031749}
\end{figure*}

KIC\,11037149 does not appear to demonstrate spot-based amplitude modulation in its light curve (Fig.~\ref{fig:11031749}), nor does it show signs of rotational frequency splitting - consistent with a lack of rotational modulation. Despite this, it is clear that it possesses unusual chemical abundances of Sr, Cr, and Eu from its spectrum (Sec.~\ref{sec:spectra}). We theorise two possibilities behind the lack of observable modulation. If the angles of inclination or magnetic obliquity are close to $0^\circ$, no modulation would be observed since the axis of rotation is pointing towards \kepler. However, this assumes that the chemical abundance spots are aligned over the magnetic poles, which has been shown to not always be the case \citep{Kochukhov2004Multielement}. Another possibility is that the period of rotation could be much longer than the time-base of the \kepler\ LC data. While the typical rotational period for A-type stars is rather short \citep{Royer2007Rotational,Royer2009Rotation,Murphy2015Observational}, the rotation for Ap types can exceed even 10 yr due to the effects of magnetic braking \citep{Landstreet2000Magnetic}. Indeed, a non-negligible fraction of Ap stars are known to have rotational periods exceeding several centuries \citep{Mathys2015Very}, and so it is possible for the star to simply be an extremely slow rotator.

The aliased signal of the true pulsation is visible even in the unfiltered LC data (Fig.~\ref{fig:11031749}). After filtering, we identified one pulsation frequency ($\nu_1$: 1372.72\,\muhz), and provide a fit in Table~\ref{tab:pulsators}. With no clear multiplet structure around the primary frequency, we are unable to constrain the inclination and magnetic obliquity within the framework of the oblique pulsator model. Indeed, without any apparent low-frequency modulation we are unable to even provide a rotation period. This represents an interesting, although not unheard of challenge for determining the rotation period. Since the photometric and spectral variability originate with the observed spot-based modulation, neither method can determine the rotation period without a longer time-base of observations. Regardless, we include KIC\,11031749 in our list of new roAp stars as it satisfies the main criterion of exhibiting both rapid oscillations and chemical abundance peculiarities.

\subsection{KIC~11296437}
\begin{figure*}
    \centering
    \includegraphics[width=\linewidth]{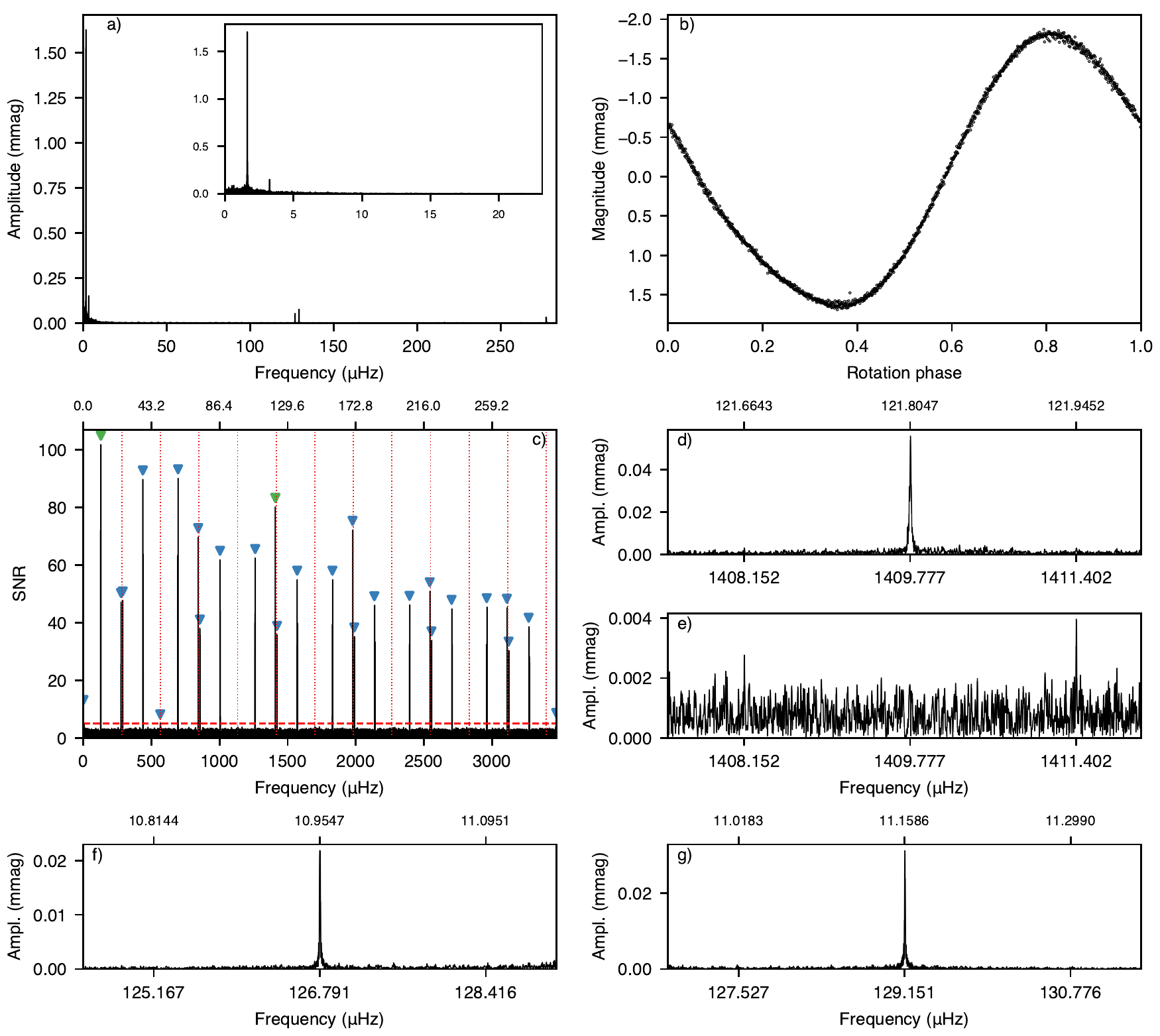}
    \caption{Same as Fig.~\ref{fig:6631188} for KIC\,11296437. In the SNR periodogram, it is not always the case that the power distribution in the super-Nyquist regime will ensure that the highest amplitude frequency will also be the highest SNR frequency.}
    \label{fig:11296437}
\end{figure*}

\begin{figure*}
    \centering
    \includegraphics[width=\linewidth]{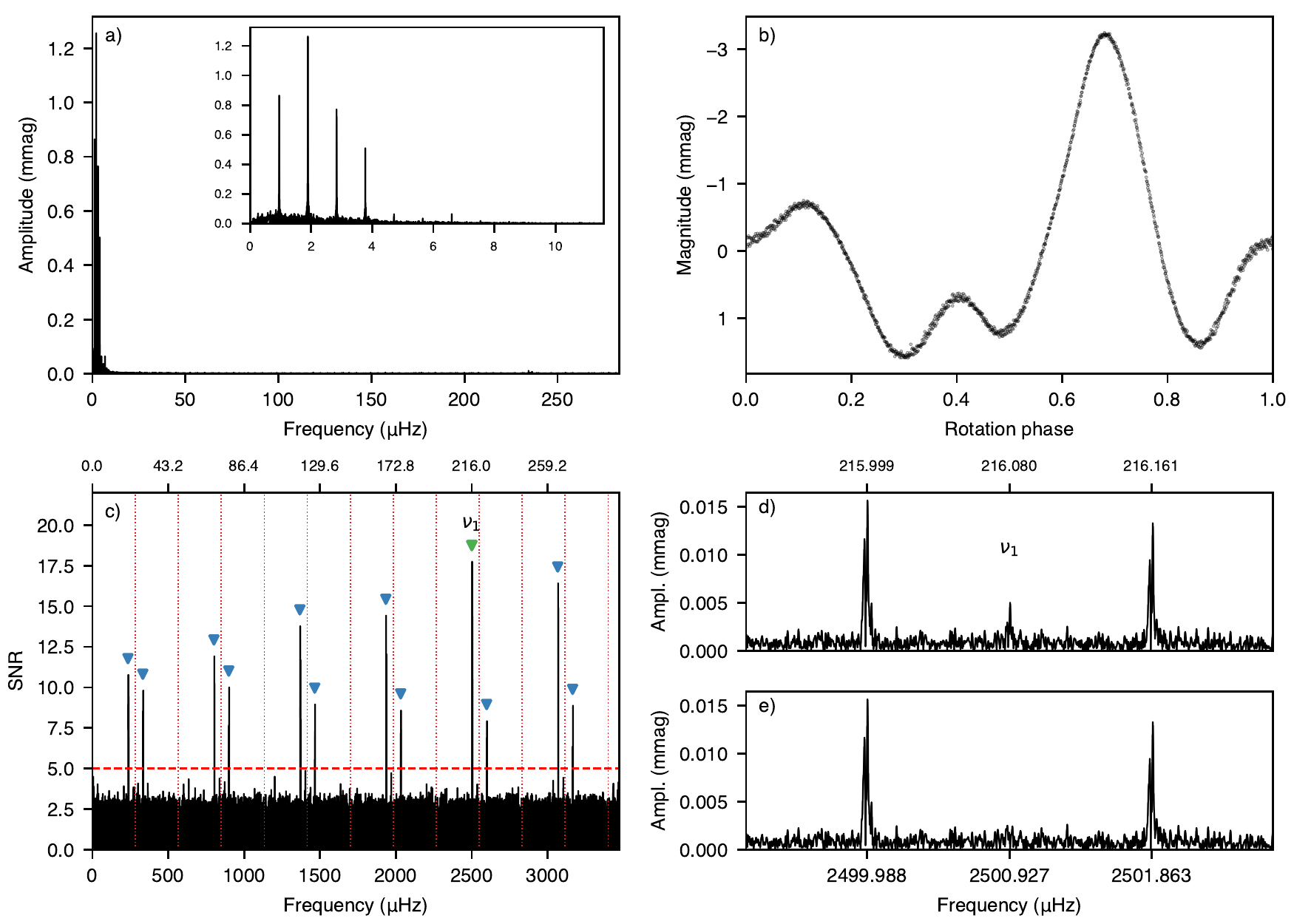}
    \caption{\textbf{a)} Amplitude spectrum of KIC\,11409673 out to the nominal Nyquist frequency. The inset shows the low-frequency region of the spectrum corresponding to the rotation frequency. \textbf{b)} Light curve folded and binned at a ratio of 50:1. \textbf{c)} Amplitude spectrum after a high-pass filter has removed the low-frequency signals -- the true oscillation frequency can be observed as the signal of maximum amplitude (green). All other peaks flagged as aliases above the 5 SNR are marked in blue. The red dashed lines denote integer multiples of the Nyquist frequency. \textbf{d)} Zoomed region of the primary frequency before, and \textbf{e)}, after pre-whitening $\nu_1$. The top x-axis, where available, is the corresponding frequency in d$^{-1}$.}
    \label{fig:11409673}
\end{figure*}

KIC\,11296437 is a known rotationally variable star \citep{Reinhold2013Rotation}, whose period of rotation at 7.12433\,$\pm$\,0.00002\,d is evident in the low-frequency region of the amplitude spectrum (Fig.~\ref{fig:11296437}). Folding the light curve on this period shows only a single spot or set of spots, implying that $i$+$\beta < 90^\circ$. 

The primary pulsation frequency was found to be 1409.78\,\muhz\ in the high-pass filtered light curve. This mode shows two sidelobes split by 1.625\,\muhz\ which is in good agreement with the rotation frequency. Two low-frequency modes, $\nu_2$ and $\nu_3$, are also present below the Nyquist frequency, at 126.79 and 129.15\,\muhz\ respectively. These modes are clearly non-aliased pulsations, as they are not split by the \kepler\ orbital period. However, neither of them are split by the rotational frequency. We provide a fit to these frequencies in Table~\ref{tab:pulsators}.

We apply the oblique pulsator model to $\nu_1$ by assuming the mode is split by the rotation frequency, and find that all three phases agree within error, implying that $\nu_1$ is not distorted. The same test can not be applied to $\nu_2$ and $\nu_3$ due to their lack of rotational splitting. We can further constrain the geometry of the star by again considering the sidelobe amplitude ratios (Eq.\,\ref{eqn:dipole}). Using the values in Table~\ref{tab:11296437_forcefit}, we find that $\tan{i}\tan{\beta} = 0.11\,\pm\,0.01$, and provide a summary of angles satisfying these values in Fig.~\ref{fig:ibeta_combo}, which demonstrates that only one pulsation pole should be observed over the rotation cycle ($i$+$\beta < 90^\circ$).

\begin{table}
    \centering
    \caption{Linear least squares fit to the pulsation and force-fitted sidelobes in KIC\,11296437. The zero-point for the fit is BJD 2455690.64114.}
    \label{tab:11296437_forcefit}
    \begin{tabular}{lccc}
        \hline
        ID    & Frequency & Amplitude$_{\rm \,\,intrinsic}$ & Phase  \\
         & (\muhz) & (mmag) & (rad) \\
        \hline
        $\nu_1 - $\nurot    & $1408.15213$ & $0.020\,\pm\,0.002$    & $-0.678\,\pm\,0.117$    \\
        $\nu_1$             & $1409.77671$  & $0.352\,\pm\,0.002$    & $-0.681\,\pm\,0.007$       \\
        $\nu_1 + $\nurot    & $1411.40129$ & $0.018\,\pm\,0.002$    & $-0.678\,\pm\,0.131$       \\
        \hline
    \end{tabular}
\end{table}

KIC\,11296437 is highly unusual in that it displays both high-frequency roAp pulsations and low-frequency $p-$mode pulsations that are typically associated with $\delta$ Scuti stars. A lack of rotational splitting in the low-frequency modes suggests that the star might be a binary composed of a $\delta$ Scuti and roAp star component. If KIC\,11296437 is truly a single component system, then it would pose a major challenge to current theoretical models of roAp stars. In particular, the low-frequency modes at 126.79 and 129.15\,\muhz\ are expected to be damped by the magnetic field according to previous theoretical modelling \citep{Saio2005Nonadiabatic}. KIC\,11296437 would be the first exception to this theory amongst the roAp stars. 

On the other hand, it would also be highly unusual if KIC\,11296437 were a binary system. It is rare for Ap stars to be observed in binaries, and much more so for roAp stars. Currently, there is one known roAp star belonging to a spectroscopic binary \citep{Hartmann2015Radialvelocity}, with several other suspected binaries \citep{Scholler2012Multiplicity}. Stellar multiplicity in roAp stars is important for understanding their evolutionary formation and whether tidal interactions may inhibit their pulsations.

\subsection{KIC~11409673}

KIC\,11409673 is a peculiar case, as it has previously been identified as an eclipsing binary, and later a heartbeat binary \citep{Kirk2016Kepler}. We note that a radial velocity survey of heartbeat stars has positively identified KIC\,11409673 as a roAp star \citep{Shporer2016Radial}. Here we provide independent confirmation of this result through super-Nyquist asteroseismology, as their result has been ignored in later catalogues of roAp stars. KIC\,11409673 has a clear low-frequency variation at 0.94\,\muhz, corresponding to a rotational period of 12.3107\,$\pm$\,0.0003 d. Similar to KIC\,6631188, the low-frequency region is dominated by a higher amplitude signal at 2$\nu_{\rm rot}$ consistent with observations of the double-wave nature, as seen in Fig.~\ref{fig:11409673}. The rotation period of 12.31\,d is found by folding the light curve, and is confirmed after high-pass filtering the light curve and examining the triplet centred around the primary frequency of 2500.93\,\muhz. Similar to KIC\,7018170, KIC\,11409673 exhibits strong frequency variation which negatively affects the shape of the multiplets. We thus split the LC data into four equally spaced sections and analysed the multiplet separately in each section. This reduced the issues arising from frequency variation, despite leading to a decrease in frequency resolution. The results of the least squares analysis for the first section of data are presented in Table \ref{tab:pulsators}.

Again, applying the oblique pulsator model by assuming that the sidelobes be separated from the primary frequency by the assumed rotation frequency, we fit each section of data via least squares. By choosing the zero-point in time such that the phases of the sidelobes are equal, we are able to show that the mode is not distorted, as the three phases agree within error. This is the case for all four separate fits. The results of this test for the first section of the data are shown in Table~\ref{tab:11409673_forcefit}, with the remaining sections in Appendix~\ref{tab11409673_pulsations_appendix}. We find that $\tan{i}\tan{\beta}=6.1\,\pm\,1.1$ using Eqn.\,\ref{eqn:dipole} for the first section of data, implying that both spots are observed in agreement with the light curve.

\begin{table}
    \centering
    \caption{Linear least squares fit to the pulsation and force-fitted sidelobes in KIC\,11409673. The zero-point for the fit is BJD 2455144.43981. The data have been split into four equally spaced sets, with the sidelobes force-fitted in each set. We show only the results of the first-set below and provide the rest in Appendix\,\ref{sec:app}. The results for each set are similar, and agree within the errors.}
    \label{tab:11409673_forcefit}
    \begin{tabular}{lccc}
        \hline
        ID    & Frequency & Amplitude$_{\rm \,\,intrinsic}$ & Phase  \\
         & (\muhz) & (mmag) & (rad) \\
        \hline
        $\nu_1 - $\nurot    & $2499.98645$ & 0.306\,$\pm$\,0.018    & $1.0173\,\pm\,0.0579$    \\
        $\nu_1$             & $2500.92665$  & 0.097\,$\pm$\,0.018    & $1.1551\,\pm\,0.1834$       \\
        $\nu_1 + $\nurot    & $2501.86684$ & 0.280\,$\pm$\,0.018    & $1.0173\,\pm\,0.0632$       \\
        \hline
    \end{tabular}
\end{table}

\section{Discussion}
\subsection{Acoustic cutoff frequencies}
As discussed in Sec.~\ref{sec:intro}, several roAp stars are known to oscillate above their theoretical acoustic cutoff frequency ($\nu_{\rm ac}$). The origin of the pulsation mechanism in these super-acoustic stars remains unknown, and presents a significant challenge to theoretical modelling. We therefore calculate whether the stars presented here oscillate above their theoretical acoustic cutoff frequency following the relation 

\begin{eqnarray}
    \dfrac{\nu_{\rm ac}}{\nu_{\rm ac, \odot}} = \dfrac{{\rm M}/{\rm M}_\odot ({\rm T}_{\rm eff}/{\rm T}_{\rm eff, \odot})^{3.5}}{{\rm L}/{\rm L}_\odot},
    \label{eqn:cutoff}
\end{eqnarray}
where $\nu_{\rm ac, \odot}$ = 5300\,\muhz\ is the theoretical acoustic cutoff frequency of the Sun \citep{Jimenez2011ACOUSTIC}. Using the values provided in Table~\ref{tab:sample}, we find that KIC\,7018170 and KIC\,11409673 both oscillate above their theoretical limit (1739.5 and 2053.7\,\muhz\ respectively). The remaining stars do not oscillate above their theoretical acoustic cutoff frequency. KIC\,11031739, however, lies almost exactly on the border of the acoustic cutoff frequency within the errors.
 
\subsection{Spectral classification}
\label{sec:spectra}
The LAMOST \citep[Large Sky Area Multi-Object Fiber Spectroscopic Telescope;][]{Zhao2012LAMOST} survey has collected low-resolution spectra between 3800-9000\,\AA\ for objects in the \textit{Kepler} field. 
We obtained LAMOST spectra from the 4th Data Release (DR4). All stars presented here have at minimum one low-resolution spectrum available from LAMOST. However, the spectrum of KIC\,6631188 is of unusable SNR. We thus obtained a high-resolution spectrum of KIC\,6631188 on April 17 2019 using the HIRES spectrograph \citep{Vogt1994HIRES} at the Keck-I 10-m telescope on Maunakea observatory, Hawai`i. The spectrum was obtained and reduced as part of the California Planet Search queue \citep[CPS,][]{Howard2010CALIFORNIA}. We obtained a 10-minute integration using the C5 decker, resulting in a S/N per pixel of 30 at $\sim$\,6000\,\AA\ with a spectral resolving power of $R\sim$\,60 000. This spectrum has been down-sampled to match the MK standard spectrum.

\begin{figure*}
    \centering
    \includegraphics[width=\linewidth]{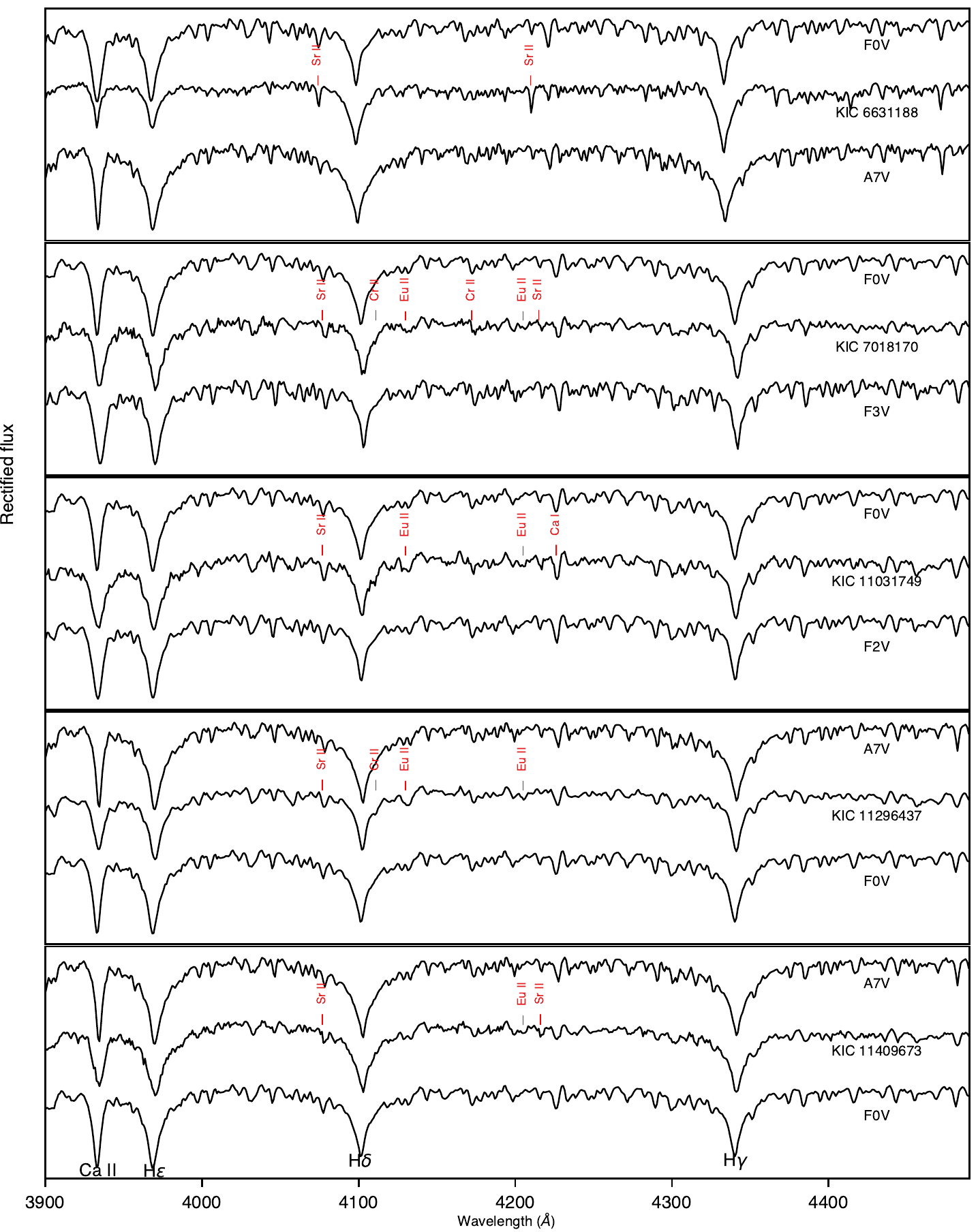}
    \caption{Keck and LAMOST spectra of KIC\,6631188, KIC\,7018170, KIC\,11031749, KIC\,11296437, and KIC\,11409673 from top to bottom. MK standard spectra have been down-sampled to match the LAMOST resolution. MK standard spectra have been obtained from \url{https://web.archive.org/web/20140717065550/http://stellar.phys.appstate.edu/Standards/std1_8.html}
    }
    \label{fig:lamost_spectra}
\end{figure*}

Fig.~\ref{fig:lamost_spectra} presents the spectra of KIC\,6631188, KIC\,7018170, KIC\,11031749, KIC\,11296437, and KIC\,11409673, with MK standard stars down-sampled to match the resolution of either the HIRES or LAMOST via the SPECTRES package \citep{Carnall2017SpectRes}. KIC\,10685175 has a pre-existing spectral classification of A4\,V\,Eu \citep{Hummerich2018Kepler}, and thus is not re-classified in this work.

In KIC 6631188, there is a strong enhancement of Sr\,{\sc{ii}} at 4077\,\AA\ and 4215\,\AA. The 4111\,\AA\ line of Cr\,{\sc{ii}} is present, which is used to confirm a Cr peculiarity, but the strongest line of Cr\,{\sc{ii}} 4172\,\AA\ line is not enhanced. The hydrogen lines look narrow for a main-sequence star, but the metal lines are well bracketed by A9\,V and F1\,V. We place this star at F0\,V\,Sr.

KIC\,7018170 shows evidence of chemical peculiarities which are only mild, but the spectrum is of low SNR. The 4077\,\AA\ line is very strong, which is indicative of an over abundance of Sr or Cr or both, but matching peculiarities in other lines of these elements are less clear. Sr\,{\sc{ii}} at 4215\,\AA\ is marginally enhanced, and the 4111 and 4172\,\AA\ lines of Cr\,{\sc{ii}} are also only marginally enhanced. The Eu\,{\sc{ii}} line at 4205\,\AA\ is significantly enhanced and is matched with a small enhancement at $4128-4132$\,\AA. We place KIC\,7018170 as a F2\,V\,(SrCr)Eu type star.

The Ca\,{\sc{ii}} K line in KIC\,11031749 is broad and a little shallow, typical of magnetic Ap stars. The 4077\,\AA\ line is very strong, suggesting enhancement of Sr and/or Cr. A mild enhancement of other Sr and Cr lines suggests both are contributing to the enhancement of the 4077\,\AA\ line. There is mild enhancement of the Eu\,{\sc ii} 4205\,\AA\ line and the 4128-4132\,\AA\ doublet suggesting Eu is overabundant. It is noteworthy that the Ca\,{\sc{i}} 4226\,\AA\ line is a little deep. We thus classify KIC\,11031749 as F1\,V\,SrCrEu.

In KIC\,11296437, there is a strong enhancement of Eu\,{\sc ii} at 4130\,\AA\ and 4205\,\AA. There is no clear enhancement of Sr\,{\sc ii} at 4216\,\AA\ but a slightly deeper line at 4077\,\AA\ which is also a line of Cr\,{\sc ii}. The 4111\,\AA\ line of Cr\,{\sc ii} is present, which is used to confirm a Cr peculiarity, but the 4172\,\AA\ line does not look enhanced which is normally the strongest line. The hydrogen lines look narrow for a main-sequence star, but the metal lines are well bracketed by A7\,V and F0\,V. We place this star at A9\,V EuCr.

For KIC\,11409673, there is enhanced absorption at 4216\,\AA\ which is a classic signature of a Sr overabundance in an Ap star. This is usually met with an enhancement in the 4077\,\AA\ line, but that line is also a line of Cr. The 4077\,\AA\ line is only moderately enhanced, but enough to support a classification of enhanced Sr. Since the 4172\,\AA\ line is normal, it appears that Cr is not enhanced. Other Cr lines cannot be relied upon at this SNR. Since the Eu\,{\sc{ii}} 4205\,\AA\ absorption line is strong, it appears that Eu is overabundant. There is no other evidence for a Si enhancement. The Ca\,{\sc{ii}} K line is a little broad and shallow for A9, which suggests mild atmospheric stratification typical of magnetic Ap stars. The hydrogen lines are a good fit intermediate to A7 and F0. We thus classify this star as A9\,V\,SrEu.

\begin{figure*}
    \centering
    \includegraphics[width=\linewidth]{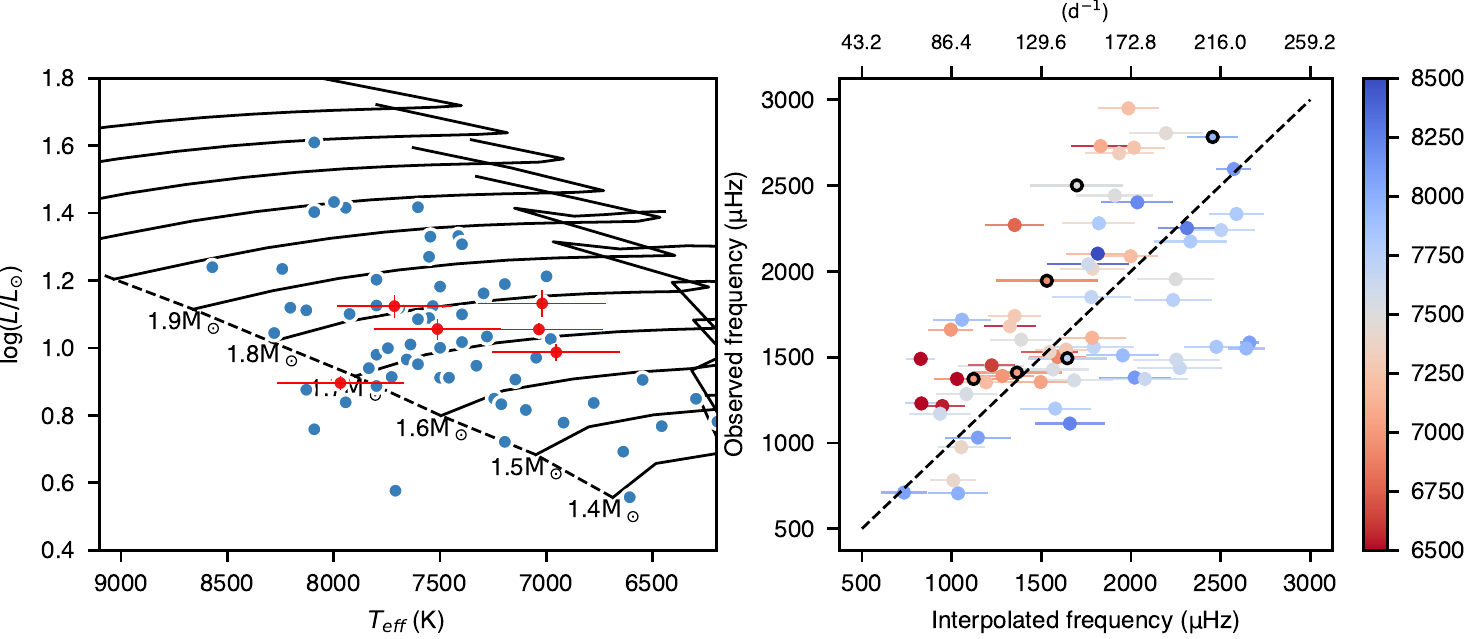}
    \caption{Left panel: Positions of the previously known (blue) and new roAp stars (red) discussed in this paper. Uncertainties have only been shown on the new roAp stars for clarity. Although stellar tracks are computed in 0.05-M$_\odot$ intervals, only every second track has been displayed here. Right panel: Interpolated and observed frequencies from pulsation modelling of the roAp stars, coloured by effective temperature. The observed frequencies are taken as the signal of highest amplitude. Circles mark previously known roAp stars, whereas outlined circles mark the six new stars. Uncertainties on the interpolated frequencies are obtained through a Monte Carlo simulation.}
    \label{fig:HRD}
\end{figure*}

\subsection{Positions in the H-R diagram}
\label{sec:modelling}

To place our sample on the HR diagram (Fig.~\ref{fig:HRD}), we derived new stellar tracks based on the models of \citet{Cunha2013Testing}. We performed linear, non-adiabatic pulsation calculations for a grid of models covering the region of the HR diagram where roAp stars are typically found. We considered models with masses between 1.4 and 2.5 M$_\odot$, in steps of 0.05 M$_\odot$, and fixed the interior chemical composition at $Y=0.278$ and $X=0.705$. 

The calculations followed closely those described for the polar models discussed in \citet{Cunha2013Testing}. The polar models consider that envelope convection is suppressed by the magnetic field, a condition required for the excitation by the opacity mechanism of high radial order modes in roAp stars. Four different cases were considered for each fixed effective temperature and luminosity in the grid. The first case considered an equilibrium model with a surface helium abundance of $Y_{\rm surf}=0.01$ and an atmosphere that extends to a minimum optical depth of $\tau_{\rm min}=3.5\times 10^{-5}$. For this case the pulsations were computed with a fully reflective boundary condition. The other three cases considered were in all similar to this one, except that the above options were modified one at a time to: $Y_{\rm surf}=0.1$; $\tau_{\rm min}=3.5\times 10^{-4}$; transmissive boundary condition \citep[see][for further details on the models]{Cunha2013Testing}. We provide a summary of these models in Table~\ref{tab:models}. We note that the impact of the choice of $Y$ and $X$ on the frequencies of the excited oscillations is negligible compared to the impact of changing the aspects of the physics described above. 

\begin{table}
	\centering
	\caption{Model parameters of the non-adiabatic calculations. Shown are the surface helium abundance $Y_{\rm surf}$, minimum optical depth $\tau_{\rm min}$, and outer boundary condition in the pulsation code.}
	\label{tab:models}
	\begin{tabular}{lccc} 
		\hline
		Model & $Y_{\rm surf}$ & $\tau_{\rm min}$ & Boundary condition \\
		\hline
		1 &$0.01$ & $3.5x10^{-5}$& Reflective\\
		2 & $0.1$ & $3.5x10^{-5}$& Reflective \\
		3 &$0.01$ & $3.5x10^{-4}$& Reflective \\
		4 &$0.01$ & $3.5x10^{-5}$& Transmissive \\
		\hline
	\end{tabular}
\end{table}

For each fixed point in the track we calculated the frequency of maximum growth rate as a comparison to our observed frequencies. The observed frequency was assumed to be the mode of highest linear growth rate. However, it should be noted that this may not necessarily be the case. Using these tracks, we performed linear interpolation to obtain an estimate of the masses and frequencies derived from modelling. Uncertainties on the masses have been artificially inflated to account for uncertainties in metallicity, temperature and luminosity (0.2, 0.1, and 0.05\,$\rm M/\rm M_\odot$ respectively), following \citet{Murphy2019Gaiaderived}. An extra error component of 0.1\,$\rm M/\rm M_\odot$ is included to account for unknown parameters in the stellar modelling, such as overshooting and mixing length. These four contributions are combined in quadrature, yielding a fixed uncertainty of 0.25\,$\rm M/\rm M_\odot$. Frequency interpolation is performed for each model (1 through 4), with the plotted value being the median of these results. The uncertainty in the interpolation of both frequency and mass is obtained from a Monte Carlo simulation sampled from the uncertainty in the temperature and luminosity of the stars. We s how the results of the positions in the H-R diagram and comparison of interpolated frequencies in Fig.~\ref{fig:HRD}.

For frequencies below $\sim$1800\,\muhz\, the agreement between theory and observations is reasonable, albeit with discrepancies on a star-by-star case that may be due to an incomplete modelling of the physics of these complex stars. However, for stars with higher characteristic frequencies there seems to be two distinct groups, one lying below and the other clearly above the 1:1 line. Coloured by temperature, we note that the group lying above the 1:1 line tends to be cooler in general. The suppression of envelope convection is key to the driving of roAp pulsations by the opacity mechanism. As it is harder for suppression to take place in the coolest evolved stars, one may question whether an additional source of driving is at play in these stars. In fact, it has been shown in a previous work that the driving of the very high frequency modes observed in some well known roAp stars cannot be attributed to the opacity mechanism. It was argued that they may, instead, be driven by the turbulent pressure if envelope convection is not fully suppressed \citep{Cunha2013Testing}. Whether that mechanism could contribute also to the driving of the modes observed in the stars laying clearly above the 1:1 line on the right panel of Fig. 10 is something that should be explored in future non-adiabatic modelling of roAp stars.

\subsection{Intrinsic amplitude and phase variability}
\label{sec:amp_variability}

\begin{figure*}
    \centering
    \includegraphics[width=\linewidth]{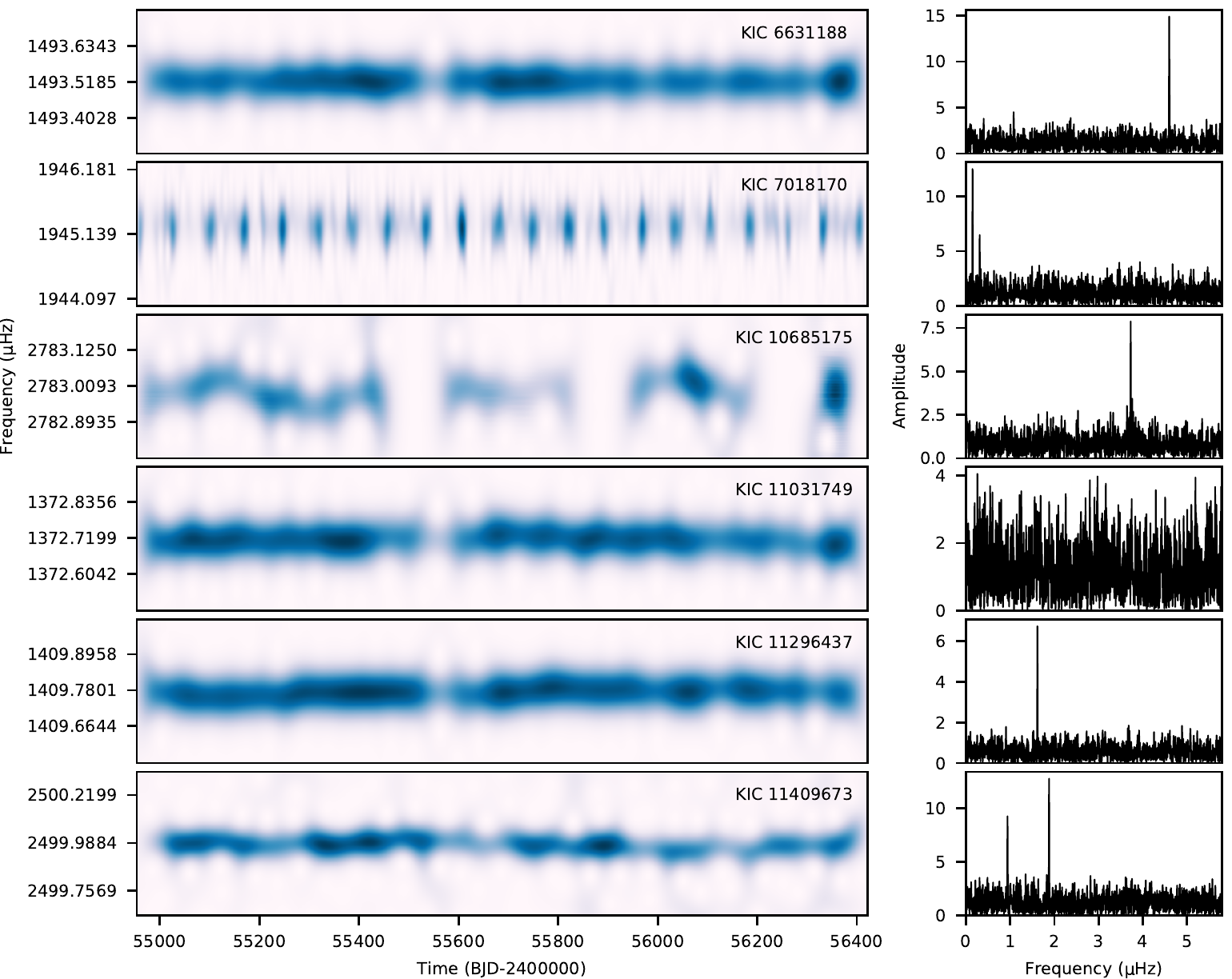}
    \caption{Time and frequency domain analysis of the primary frequencies for each of the roAp stars, where the colour shows the normalised amplitude of the signal. The amplitude spectra (right) are obtained by taking the periodogram of the wavelet at the primary frequency, and are arbitrarily normalised. All but KIC\,11031749 show strong frequencies in agreement the spot-based rotational modulation of the light curve, confirming their oblique pulsating nature. The wavelet for KIC\,11409673 is taken from the sidelobe frequency $\nu_1$-\nurot\ due to the central frequency being of low amplitude. The gaps in KIC\,10685175 are due to missing quarters in the \kepler\ photometry.}
    \label{fig:modulation}
\end{figure*}

\begin{figure}
    \centering
    \includegraphics[width=\linewidth]{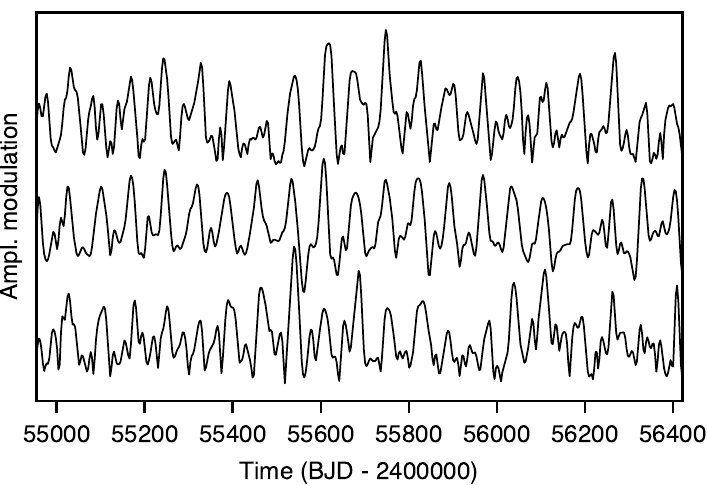}
    \caption{Amplitude modulation of the modes $\nu_1$, $\nu_2$, and $\nu_3$ in KIC\,7018170 from top to bottom. All three modes are modulated in phase with each-other, in agreement with the oblique pulsator model.}
    \label{fig:701_all_mod}
\end{figure}

\begin{figure*}
    \centering
    \includegraphics[width=\linewidth]{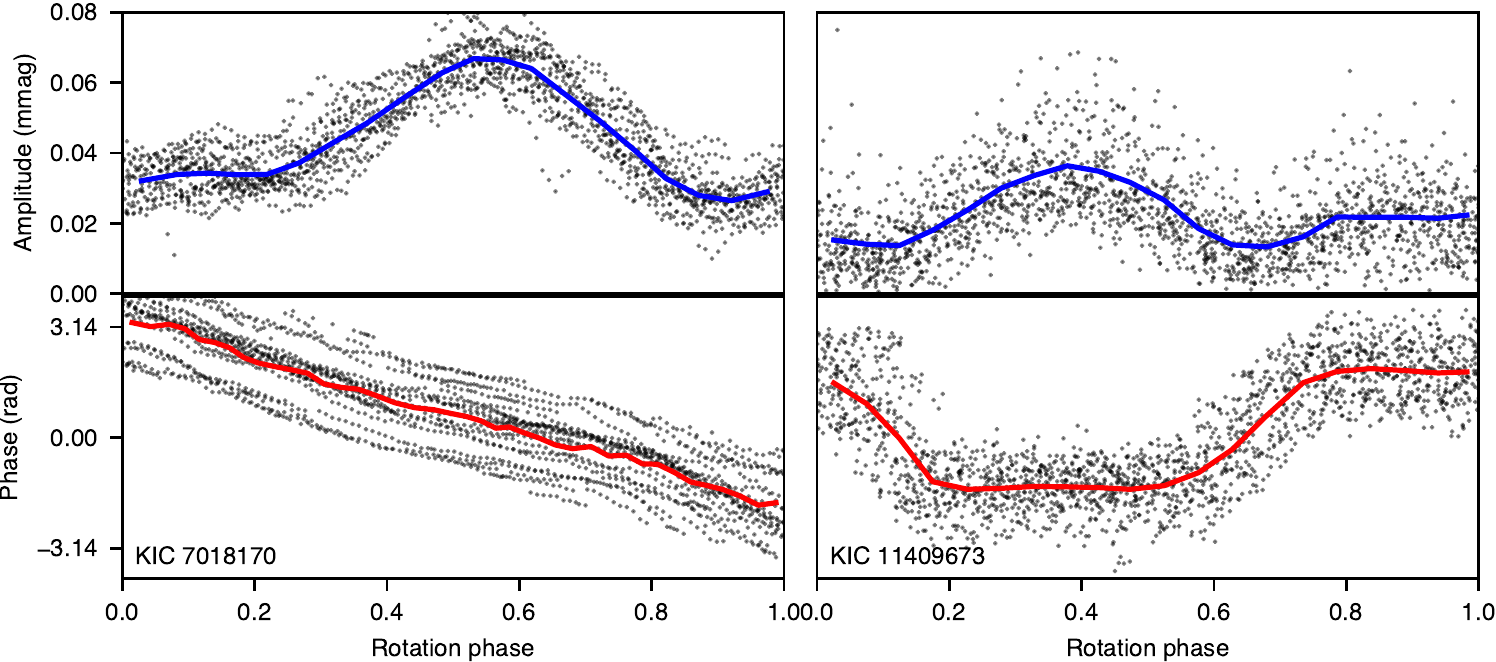}
    \caption{Amplitude and phase variations of KIC\,7018170 (left) and KIC\,11409673 (right). The top panels show the folded amplitude variations on the rotation period, whereas the bottom panels show the folded phase variations. The pulsation amplitude coincides with the rotational light extremum. A $\pi$ rad phase change is observed when the line-of-sight passes over pulsation nodes. For KIC\,7018170, the phase change occurs at a rotational phase of 0.} The solid lines mark the binned data for clarity.
    \label{fig:11409673_crossing}
\end{figure*}

Many of the known roAp stars have shown significant variation in the amplitudes and phases of their pulsation frequencies over the observation period. To examine this variability in our sample, we conducted a time and frequency domain analysis where a continuous amplitude spectrum was generated by sliding a rectangular window across the light curve. The Fourier amplitude and phase was then calculated within the window at each point. For each star, the window length was chosen to be $100/\nu_1$\,d in width to minimise phase and amplitude uncertainty whilst correctly sampling the frequency. Both rectangular and Gaussian windows were tested and found to have minimal difference in the resultant amplitudes. Before calculating the variability, the rotational frequency and corresponding harmonics were pre-whitened manually to minimise any potential frequency beating. 

Amplitude modulation of the normal modes provides a measure of the rotation period of these stars and confirms the nature of their oblique pulsations. If the pulsation amplitudes goes to zero, then a node crosses the line-of-sight and is observed, and (for a dipole mode) the pole is at $90^\circ$ to the line-of-sight. However, if the geometry is such that the amplitude never goes to zero (e.g. $\alpha$~Cir), then we must always see a pole. This leads to a periodic modulation of the pulsation amplitude, which is different to the spot-based rotational modulation seen in the light curves of Ap stars in general. Similarly, a $\pi$\,rad phase change should be observed in the phase of the frequency whenever a node crosses the line-of-sight. The continuous amplitude spectrum and corresponding rotational signal is shown in Fig.~\ref{fig:modulation}. The rotation signal was  obtained by examining the amplitude spectrum of the modulation along the primary frequency of each star.

KIC\,6631188 has obvious low-frequency modulation in its light curve, making identification of the rotation period straightforward (Sec.~\ref{sec:pulsations}). Regardless, it makes for a useful test case for confirming modulation of its primary frequency. The amplitude spectrum of the modulation signal in Fig.~\ref{fig:modulation} has a curious peak at twice the rotational frequency, 4.60\,\muhz, corresponding to a period of 2.52\,d. This result implies that the spots are not necessarily aligned along the magnetic and pulsation axes, as was shown to be possible by \cite{Kochukhov2004Multielement}.

The analysis of KIC\,7018170 greatly benefits from amplitude modulation of its modes, as the \mbox{PDCSAP} flux pipeline removes any low-frequency content in the light curve. The modulation is calculated for all three modes present in the LC data, with the amplitude spectrum taken on the weighted average signal, which is thus dominated by $\nu_1$. The frequency components of each mode multiplet are in phase, in good agreement with the oblique pulsator model. Indeed, it is quite remarkable that the secondary modes ($\nu_2$, $\nu_3$) in KIC\,7018170 exhibit such clear amplitude modulation despite being of low SNR (Fig.~\ref{fig:701_all_mod}). The amplitude spectrum of the variation in the signal is found to peak at 0.160\,\muhz\, agreeing with the rotation frequency identified from sidelobe splitting.

No evidence of amplitude modulation can be found in KIC\,11031749. We can however speculate on the nature of the lack of amplitude modulation and attribute it to two possibilities; either the position of the pulsation pole does not move relative to the observer, or the rotation period is much longer than the 4-yr \kepler\ data. Thus, no rotation period can be ascribed based on modulation of the principal frequency. KIC\,11296437 shows a periodic signal in its amplitude modulation corresponding to the rotational period for only the high-frequency ($\nu_1$) mode, with a frequency of 1.63\,\muhz. The low-frequency modes ($\nu_2$, $\nu_3$) show no evidence of amplitude modulation, suggesting that they could possibly belong to an orbital companion. KIC\,11409673 has a clear low-frequency signal and harmonic present in the light curve beginning at 0.940\,\muhz. Amplitude modulation of its primary oscillation frequency shows evidence of rotation in good agreement with the low-frequency signal. The peak in the amplitude spectrum of the modulation confirms the rotation period of 12.310\,d derived in Sec.~\ref{sec:pulsations}.

Fig.~\ref{fig:11409673_crossing} shows the variation of the pulsation amplitude and phase over the rotation period of the two stars in this work found to have observable phase crossings, KIC\,7018170 and KIC\,11409673. The maximum amplitude coincides with light maximum, which is expected when the spots producing the light variations are closely aligned with the magnetic and pulsation poles. The amplitude does not reach zero in KIC\,7018170, but almost does in KIC\,11409673. Since the amplitude does not go to zero for KIC\,7018170, we can see that the mode is distorted.

\subsection{Phase modulation from binarity}

\begin{figure*}
    \centering
    \includegraphics[width=\linewidth]{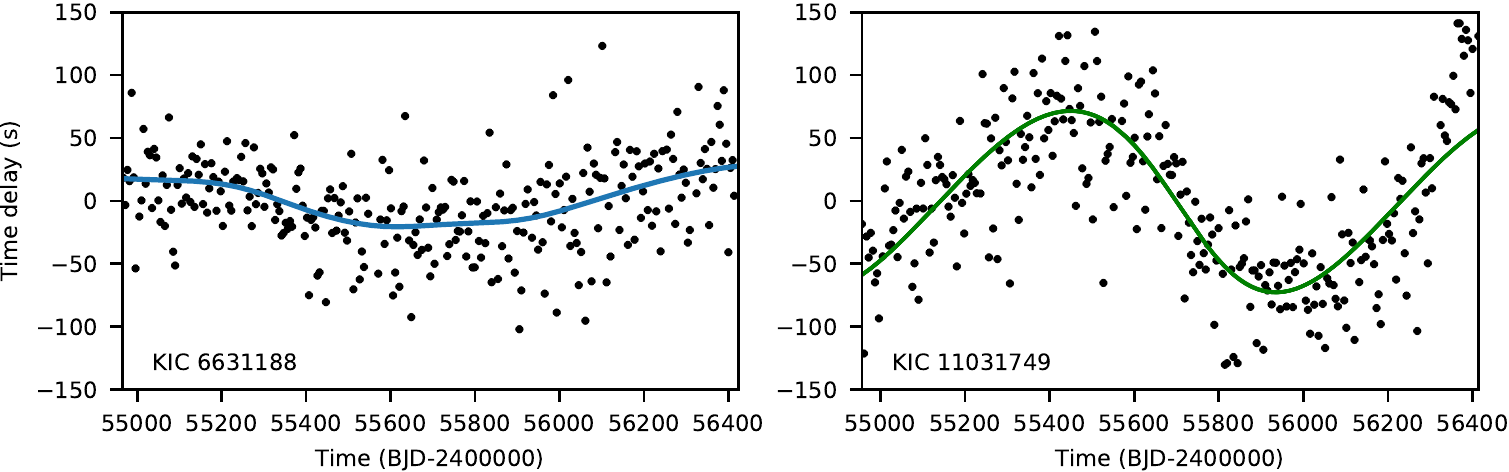}
    \caption{Left panel: Observed time delay for KIC\,6631188. The time delay, $\tau$, is defined such that it is negative when the pulsating star is nearer to us than the barycenter of the system. The blue line is not a orbital solution fit, but rather the time delay as obtained by binning the signal for clarity. If the observed signal is truly periodic, it appears to be on a timescale longer than the LC data. Right panel: Observed time delays (black dots) and the fitted model (green line) for KIC\,11031749. The phase modulation suggests a binary system of low eccentricity, whose orbital period is comparable to or longer than the time-base of the LC photometry.}
    \label{fig:11031749_PM}
\end{figure*}

Both KIC\,6631188 and KIC\,11031749 show long term phase variations independent of their rotation. If the frequency variability of KIC\,11031749 is modelled as modulation of its phase due to binary reflex motion from an orbital companion, its orbital properties can be derived following the phase modulation technique \citep{Murphy2014Finding}. To examine this, the light curve was separated into 5-day segments, where the phase of each segment at the primary frequency was calculated through a discrete Fourier transform. We then converted the phases into light arrival times ($\tau$) by dividing by the frequency, from which a map of the binary orbit was constructed. A Hamiltonian Markov Chain Monte Carlo sampler was then applied to fit the time delay curve through the use of {\sc PyMC3} \citep{Salvatier2016Probabilistic}. The sampler was run simultaneously over 4 chains for 5000 draws each, with 2000 tuning steps. The resulting fit is shown in Fig.~\ref{fig:11031749_PM}, with extracted orbital parameters in Table~\ref{tab:11031749_PM}. These parameters give a binary mass function $f(m_1, m_2, sini)$ =  0.000\,327 M$_\odot$. Assuming a primary mass $M_1$ from Table~\ref{tab:sample} of 1.78 M$_\odot$, we obtain a lower limit on the mass of the companion to be 0.105 $M_\odot$, placing its potential companion as a low-mass M-dwarf.

\begin{table}
	\centering
	\caption{Orbital parameters of KIC\,11031749 obtained through phase modulation. $\varpi$ is the angle from the nodal point to the periapsis, $i$ is the inclination angle. $a$ is the semimajor axis, $e$ is the eccentricity, and $\phi_p$ is the phase of periastron.}
	\label{tab:11031749_PM}
	\begin{tabular}{lcr} 
		\hline
		Quantity & Value & Units\\
		\hline
		$P_{\rm orb}$       & 1035.5\,$\pm$\,0.5 & d\\
		$(a_1 \sin{i})/c$   & 68.86\,$\pm$\,0.13 & s\\
		$e$                 & 0.203\,$\pm$\,0.004 &  \\
		$\phi_p$            & 0.80\,$\pm$\,0.03 &  \\
		$\varpi$            & 0.08\,$\pm$\,0.01 & rad \\
		$f(m_1, m_2, sini)$ &  0.000\,327 & M$_\odot$ \\
		$M_2 \sin{i}$       & 0.105\,$\pm$\,0.001 & M$_\odot$ \\
		\hline
	\end{tabular}
\end{table}

KIC\,6631188 also shows signs of frequency modulation (Fig.~\ref{fig:11031749_PM}). However, if this modulation is truly from a stellar companion then its orbital period must be much longer than the time base of the \kepler\ data. The PM method can only be used when at least one full binary orbit is observed in the time delay curve. Thus, no orbital solution can be presented here.

It is important to note the scarcity of Ap stars in binary systems. Indeed, the much smaller subset of roAp stars have a low chance of being found in a binary \citep{Abt1973Binary, North1998Binaries,Folsom2014Candidate}, however, few techniques can adequately observe the low-mass companion presented here. Although frequency modulation in roAp stars has been inferred in the past to be a consequence of binary motion, two out of the six stars presented in this work show evidence of coherent frequency/phase modulation. Whether this modulation is a consequence of changes in the pulsation cavity, magnetic field, or externally caused by orbital perturbations of a companion remains to be seen, and requires spectroscopic follow-up to rule out orbital motion via a radial velocity analysis.

\section{Conclusions}
We presented the results of a search for rapid oscillators in the \textit{Kepler} long-cadence data using super-Nyquist asteroseismology to reliably distinguish between real and aliased pulsation frequencies. We selected over 69\,000 stars whose temperatures lie within the known range of roAp stars, and based on a search for high-frequency non-alias pulsations, have detected unambiguous oscillations in six stars - KIC\,6631188, KIC\,7018170, KIC\,10685175, KIC\,11031749, KIC\,11296437, and KIC\,11409673. LAMOST or Keck spectra of five of these stars shows that they exhibit unusual abundances of rare earth elements, the signature of an Ap star, with the final target , KIC\,10685175, already being confirmed as chemically peculiar in the literature.

This research marks a significant step in our search for roAp stars, and indeed, all high-frequency pulsators. To the best of our knowledge, this is the first time super-Nyquist asteroseismology has been used solely for identification of oscillation modes to such a high frequency. Although we expect many new roAp stars to be found in the \textit{TESS} Data Releases, \kepler\ had the advantage of being able to observe stars of much fainter magnitude for a longer time-span, revealing pulsations of lower amplitude. 

\section*{Acknowledgements}
We are thankful to the entire \textit{Kepler} team for such incredible data. DRH gratefully acknowledges the support of the Australian Government Research Training Program (AGRTP) and University of Sydney Merit Award scholarships. This research has been supported by the Australian Government through the Australian Research Council DECRA grant number DE180101104. DLH and DWK acknowledge financial support from the Science and Technology Facilities Council (STFC) via grant ST/M000877/1. MC is supported in the form of work contract funded by national funds through Fundação para a Ciência e Tecnologia (FCT) and acknowledges the supported by FCT through national funds and by FEDER through COMPETE2020 by these grants: UID/FIS/04434/2019, PTDC/FIS-AST/30389/2017 \& POCI-01-0145-FEDER-030389. DH acknowledges support by the National Science Foundation (AST-1717000).

This work has made use of data from the European Space Agency (ESA) mission {\it Gaia} (\url{https://www.cosmos.esa.int/gaia}), processed by the {\it Gaia} Data Processing and Analysis Consortium (DPAC, \url{https://www.cosmos.esa.int/web/gaia/dpac/consortium}). Funding for the DPAC has been provided by national institutions, in particular the institutions participating in the {\it Gaia} Multilateral Agreement. 

The authors wish to recognize and acknowledge the very significant cultural role and reverence that the summit of Maunakea has always had within the indigenous Hawai`ian community.  We are most fortunate to have the opportunity to conduct observations from this mountain. This research was partially conducted during the Exostar19 program at the Kavli Institute for Theoretical Physics at UC Santa Barbara, which was supported in part by the National Science Foundation under Grant No. NSF PHY-1748958

Guoshoujing Telescope (the Large Sky Area Multi-Object Fiber Spectroscopic Telescope; LAMOST) is a National Major Scientific Project built by the Chinese Academy of Sciences. Funding for the project has been provided by the National Development and Reform Commission. LAMOST is operated and managed by the National Astronomical Observatories, Chinese Academy of Sciences.




\bibliographystyle{mnras}
\bibliography{Library.bib} 



\appendix

\section{least squares fit of pulsation modes for KIC~11409673}
\label{sec:app}
We provide here the force-fitted sidelobes for the other sections of data in KIC\,11409673.
\begin{table}
    \centering
    \caption{Force-fitted pulsations in KIC\,11409673 for sections 2 through 4. The zero-points were chosen to force the first set of sidelobes to be equal, and are BJD 2455501.39754, BJD 2455870.64075, and BJD 2456240.07652 respectively.}
    \begin{tabular}{lcccr}
        \hline
        ID    & Frequency & Amplitude$_{\rm \,\,Intrinsic}$ & Phase  \\
         & (\muhz) & (mmag) & (rad) \\
        \hline 
        \textit{Section 2} \\
        $\nu_1 - $\nurot    & 2499.99203 & 0.303\,$\pm$\,0.018    & -2.239\,$\pm$\,0.058    \\
        $\nu_1$             & 2500.93222  & 0.086\,$\pm$\,0.018    & -2.437\,$\pm$\,0.206       \\
        $\nu_1 + $\nurot    & 2501.87242 & 0.255\,$\pm$\,0.018    & -2.239\,$\pm$\,0.069       \\
        \hline 
        \textit{Section 3} \\
        $\nu_1 - $\nurot    & 2499.97449 & 0.311\,$\pm$\,0.017    & -1.858\,$\pm$\,0.056    \\
        $\nu_1$             & 2500.91469  & 0.090\,$\pm$\,0.017    & -1.787\,$\pm$\,0.194       \\
        $\nu_1 + $\nurot    & 2501.85489 & 0.221\,$\pm$\,0.017    & -1.858\,$\pm$\,0.079       \\
        \hline 
        \textit{Section 4} \\
        $\nu_1 - $\nurot    & 2499.96384 & 0.281\,$\pm$\,0.017    & -2.169\,$\pm$\,0.061    \\
        $\nu_1$             & 2500.90404  & 0.079\,$\pm$\,0.017    & -2.243\,$\pm$\,0.217       \\
        $\nu_1 + $\nurot    & 2501.84424 & 0.241\,$\pm$\,0.017    & -2.169\,$\pm$\,0.071       \\
        \hline
    \end{tabular}
    \label{tab11409673_pulsations_appendix}
\end{table}


\bsp	
\label{lastpage}
\end{document}